\documentclass[apj]{emulateapj}

\usepackage{apjfonts}
\usepackage{ifthen}
\usepackage{natbib}
\usepackage{amssymb, amsmath}
\usepackage{appendix}
\usepackage{etoolbox}
\usepackage{longtable}
\usepackage{tablefootnote}
\bibpunct[, ]{(}{)}{,}{a}{}{,}

\newcommand\submitms{y}  % set to y to follow AAS ``ms'' names, etc.
\newcommand\degree{\degr}

\newcommand\vs{\emph{vs.}}

\usepackage[%pdftex,      %%% hyper-references for pdflatex
bookmarks=true,           %%% generate bookmarks ...
bookmarksnumbered=true,   %%% ... with numbers
colorlinks=true,          % links are colored
citecolor=blue,           % green   % color of cite links
linkcolor=blue,           %cyan,         % color of hyperref links
menucolor=blue,           % color of Acrobat Reader menu buttons
urlcolor=blue,            % color of page of \url{...}
linkbordercolor={0 0 1},  %%% blue frames around links
pdfborder={0 0 1},
frenchlinks=true]{hyperref}

\providecommand{\adsurl}[1]{\href{#1}{ADS}}

\makeatletter
\patchcmd{\NAT@citex}
  {\@citea\NAT@hyper@{%
     \NAT@nmfmt{\NAT@nm}%
     \hyper@natlinkbreak{\NAT@aysep\NAT@spacechar}{\@citeb\@extra@b@citeb}%
     \NAT@date}}
  {\@citea\NAT@nmfmt{\NAT@nm}%
   \NAT@aysep\NAT@spacechar\NAT@hyper@{\NAT@date}}{}{}

\patchcmd{\NAT@citex}
  {\@citea\NAT@hyper@{%
     \NAT@nmfmt{\NAT@nm}%
     \hyper@natlinkbreak{\NAT@spacechar\NAT@@open\if*#1*\else#1\NAT@spacechar\fi}%
       {\@citeb\@extra@b@citeb}%
     \NAT@date}}
  {\@citea\NAT@nmfmt{\NAT@nm}%
   \NAT@spacechar\NAT@@open\if*#1*\else#1\NAT@spacechar\fi\NAT@hyper@{\NAT@date}}
  {}{}
\makeatother

\makeatletter
\DeclareRobustCommand{\lowcase}[1]{\@lowcase#1\@nil}
\def\@lowcase#1\@nil{\if\relax#1\relax\else\MakeLowercase{#1}\fi}
\makeatother

\newcommand\tnm[1]{\tablenotemark{#1}}
\newcommand\SST{{\em Spitzer Space Telescope}}
\newcommand\Spitzer{{\em Spitzer}}
\newcommand\chisq{$\chi\sp2$}

\newcommand\ms{m\,s$\sp{-1}$}

\newcommand\mjup{$M\sb{\rm Jup}$}
\newcommand\rjup{$R\sb{\rm Jup}$}
\newcommand\msun{$M\sb{\odot}$}
\newcommand\rsun{$R\sb{\odot}$}

\DeclareSymbolFont{UPM}{U}{eur}{m}{n}
\DeclareMathSymbol{\umu}{0}{UPM}{"16}
\let\oldumu=\umu
\renewcommand\umu{\ifmmode\oldumu\else\math{\oldumu}\fi}
\newcommand\micro{\umu}
\renewcommand\micron{\micro m}
\newcommand\microns{\micron}

\let\oldsim=\sim
\renewcommand\sim{\ifmmode\oldsim\else\math{\oldsim}\fi}
\let\oldpm=\pm
\renewcommand\pm{\ifmmode\oldpm\else\math{\oldpm}\fi}
\newcommand\by{\ifmmode\times\else\math{\times}\fi}
\newcommand\ttt[1]{10\sp{#1}}
\newcommand\tttt[1]{\by\ttt{#1}}
\newcommand\tablebox[1]{\begin{tabular}[t]{@{}l@{}}#1\end{tabular}}
\newbox{\wdbox}
\renewcommand\c{\setbox\wdbox=\hbox{,}\hspace{\wd\wdbox}}
\renewcommand\i{\setbox\wdbox=\hbox{i}\hspace{\wd\wdbox}}
\newcommand\n{\hspace{0.5em}}

\newcount\timect
\newcount\hourct
\newcount\minct
\newcommand\now{\timect=\time \divide\timect by 60
         \hourct=\timect \multiply\hourct by 60
         \minct=\time \advance\minct by -\hourct
         \number\timect:\ifnum \minct < 10 0\fi\number\minct}

\newcommand\mctc{\multicolumn{2}{c}}

\catcode`@=11

\newcommand\comment[1]{}
\newcommand\commenton{\catcode`\%=14}
\newcommand\commentoff{\catcode`\%=12}

\renewcommand\math[1]{$#1$}
\newcommand\mathshifton{\catcode`\$=3}
\newcommand\mathshiftoff{\catcode`\$=12}

\comment{alignment tab}
\let\atab=&
\newcommand\atabon{\catcode`\&=4}
\newcommand\ataboff{\catcode`\&=12}

\let\oldmsp=\sp
\let\oldmsb=\sb
\def\sp#1{\ifmmode
           \oldmsp{#1}%
         \else\strut\raise.85ex\hbox{\scriptsize #1}\fi}
\def\sb#1{\ifmmode
           \oldmsb{#1}%
         \else\strut\raise-.54ex\hbox{\scriptsize #1}\fi}
\newbox\@sp
\newbox\@sb
\def\sbp#1#2{\ifmmode%
           \oldmsb{#1}\oldmsp{#2}%
         \else
           \setbox\@sb=\hbox{\sb{#1}}%
           \setbox\@sp=\hbox{\sp{#2}}%
           \rlap{\copy\@sb}\copy\@sp
           \ifdim \wd\@sb >\wd\@sp
             \hskip -\wd\@sp \hskip \wd\@sb
           \fi
        \fi}
\def\msp#1{\ifmmode
           \oldmsp{#1}
         \else \math{\oldmsp{#1}}\fi}
\def\msb#1{\ifmmode
           \oldmsb{#1}
         \else \math{\oldmsb{#1}}\fi}
\def\supon{\catcode`\^=7}
\def\supoff{\catcode`\^=12}
\def\subon{\catcode`\_=8}
\def\suboff{\catcode`\_=12}
\def\supsubon{\supon \subon}
\def\supsuboff{\supoff \suboff}

\newcommand\actcharon{\catcode`\~=13}
\newcommand\actcharoff{\catcode`\~=12}

\newcommand\paramon{\catcode`\#=6}
\newcommand\paramoff{\catcode`\#=12}

\comment{And now to turn us totally on and off...}

\newcommand\reservedcharson{ \commenton  \mathshifton  \atabon  \supsubon 
                             \actcharon  \paramon}

\newcommand\reservedcharsoff{\commentoff \mathshiftoff \ataboff \supsuboff 
                             \actcharoff \paramoff}

\catcode`@=12
\reservedcharsoff

\reservedcharson

\comment{ Must have ONLY ONE of these... trust these macros, they work

}
\reservedcharsoff

\actcharon
\if\submitms y

\else

\comment{
\received{}
\revised{}
\accepted{}
}
\fi

\if\submitms y
\slugcomment{\tablebox{In preparation for {\em ApJ}. DRAFT of {\today}.}}
\else
\slugcomment{          In preparation for {\em ApJ}.  DRAFT of {\today} \now.}
\fi
\reservedcharson

\shorttitle{Multi-band Analysis of TrES-1}
\shortauthors{Cubillos {\em et al.}}

\begin{document}

\title{A {\em Spitzer} Five-Band Analysis of the Jupiter-Sized Planet
  T\MakeLowercase{r}ES-1}

\author{Patricio~Cubillos\altaffilmark{1,2},
Joseph~Harrington\altaffilmark{1,2},
Nikku~Madhusudhan\altaffilmark{3},
Andrew~S.~D.~Foster\altaffilmark{1},
Nate~B.~Lust\altaffilmark{1},
Ryan~A.~Hardy\altaffilmark{1},
and M.~Oliver~Bowman\altaffilmark{1}
}

\affil{\sp{1} Planetary Sciences Group, Department of Physics, 
       University of Central Florida, Orlando, FL 32816-2385 USA}
\affil{\sp{2} Max-Plank-Institut f\"ur Astronomie, K\"onigstuhl 17,
  D-69117, Heidelberg, Germany}
\affil{\sp{3} Department of Physics and Department of Astronomy, Yale
  University, New Haven, CT 06511, USA}
\email{pcubillos@fulbrightmail.org}

\begin{abstract}
  With an equilibrium temperature of 1200~K, TrES-1 is one of the
  coolest hot Jupiters observed by {\Spitzer}.  It was also the first
  planet discovered by any transit survey and one of the first
  exoplanets from which thermal emission was directly observed.  We
  analyzed all {\Spitzer} eclipse and transit data for TrES-1 and
  obtained its eclipse depths and brightness temperatures in the 3.6
  {\micron} (0.083\% {\pm} 0.024\%, 1270 {\pm} 110 K), 4.5 {\micron}
  (0.094\% {\pm} 0.024\%, 1126 {\pm} 90 K), 5.8 {\micron} (0.162\%
  {\pm} 0.042\%, 1205 {\pm} 130 K), 8.0 {\micron} (0.213\% {\pm}
  0.042\%, 1190 {\pm} 130 K), and 16 {\micron} (0.33\% {\pm} 0.12\%,
  1270 {\pm} 310 K) bands.  The eclipse depths can be explained,
  within 1$\sigma$ errors, by a standard atmospheric model with solar
  abundance composition in chemical equilibrium, with or without a
  thermal inversion.  The combined analysis of the transit, eclipse,
  and radial-velocity ephemerides gives an eccentricity $e =
  0.033\sp{+0.015}\sb{-0.031}$, consistent with a circular orbit.
  Since TrES-1's eclipses have low signal-to-noise ratios, we
  implemented optimal photometry and differential-evolution
  Markov-chain Monte Carlo (MCMC) algorithms in our Photometry for
  Orbits, Eclipses, and Transits (POET) pipeline.  Benefits include
  higher photometric precision and \sim10 times faster MCMC
  convergence, with better exploration of the phase space and no
  manual parameter tuning.
\end{abstract}

\section{INTRODUCTION}
\label{introduction}

Transiting exoplanets offer the valuable chance to measure the light
emitted from the planet directly.  In the infrared, the eclipse depth
of an occultation light curve (when the planet passes behind its host
star) constrains the thermal emission from the planet.  Furthermore,
multiple-band detections allow us to characterize the atmosphere of
the planet \citep[e.g.,][]{SeagerDeming2010AnnualRev}.  Since the
first detections of exoplanet occultations---TrES-1
\citep{CharbonneauEtal2005apjTrES1} and HD\,298458b
\citep{Deming2005Nat}---there have been several dozen occultations
observed.  However, to detect an occultation requires an exhaustive
data analysis, since the the planet-to-star flux ratios typically lie
below $\ttt{-3}$.  For example, for the {\SST}, these flux ratios are
lower than the instrument's photometric stability criteria
\citep{FazioEtal2004apjsIRAC}.  In this paper we analyze {\Spitzer}
follow-up observations of TrES-1, highlighting improvement in
light-curve data analysis over the past decade.

TrES-1 was the first exoplanet discovered by a wide-field transit
survey \citep{AlonsoEtal2004apjTrES1disc}.
% The Star:
Its host is a typical K0 thin-disk star
\citep{SantosEtal2006TrES1chemAbundances} with solar metallicity
\citep{LaughlinEtal2005TrES1followup, SantosEtal2006TrES1spectroscopy,
  SozzettiEtal2006TrES1starChemComp}, effective temperature $T\sb{\rm
  eff} = 5230 \pm 50$ K, mass $M\sb{*} = 0.878 \pm 0.040$ solar masses
({\msun}), and radius $R\sb{*} = 0.807 \pm 0.017$ solar radii
({\rsun}, \citealp{TorresEtal2008reanalyses}).
\citet{SteffenAgol2005TrES1transitTimes} dismissed additional
companions (with $M > M\sb{\oplus}$).
% Secondary eclipses:
\citet{CharbonneauEtal2005apjTrES1} detected the secondary eclipse in
the 4.5 and 8.0 {\micron} {\Spitzer} bands.
\citet{KnutsonEtal2007TrES1GroundThEmission} attempted ground-based
eclipse observations in the L band (2.9 to 4.3 {\microns}), but
did not detect the eclipse.

% Transit analyses:
The TrES-1 system has been repeatedly observed during transit from
ground-based telescopes \citep{NaritaEtal2007TrES1RLmeasurements,
  RaetzEtal2009TrES1Transits, Vanko2009TrES1Transits,
  RabusEtal2009TrEStransits, HrudkovaEtal2009TTVsearch,
  SadaEtal2012Transits} and from the {\em Hubble Space Telescope}
\citep{CharbonneauEtal2007TransitsReview}.  The analyses of the
cumulative data \citep{ButlerEtal2006Catalog,
  Southworth2008HomogeneousStudyI, Southworth2009HomogeneousStudyII,
  TorresEtal2008reanalyses} agree (within error bars) that the planet
has a mass of $M\sb{p} = 0.752 \pm 0.047$ Jupiter masses ({\mjup}), a
radius $R\sb{p} = 1.067 \pm 0.022$ Jupiter radii ({\rjup}), and a
circular, 3.03 day orbit, whereas \citet{WinnEtal2007apjTres1}
provided accurate details of the transit light-curve shape.  Recently,
an adaptive-optics imaging survey
\citep{AdamsEtal2013AOImagingCompanions} revealed that TrES-1 has a
faint background stellar companion ($\Delta$mag = 7.68 in the Ks band,
or 0.08\% of the host's flux) separated by 2.31{\arcsec} (1.9 and 1.3
{\Spitzer} pixels at 3.6--8~{\microns} and at 16 {\microns},
repectively).  The companion's type is unknown.

\begin{figure*}[t]
\centering
\includegraphics[width=0.49\linewidth, clip]{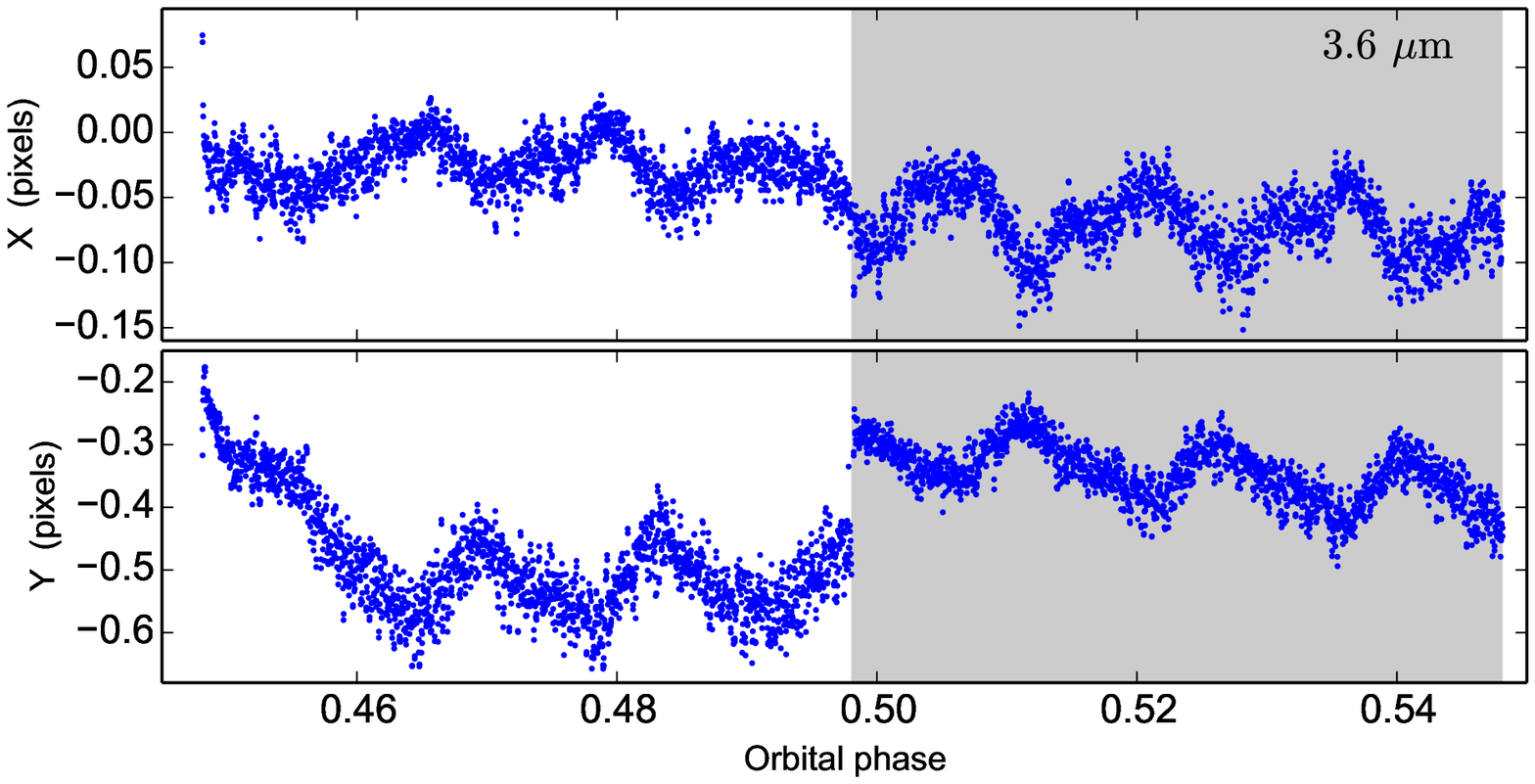}\hfill
\includegraphics[width=0.49\linewidth, clip]{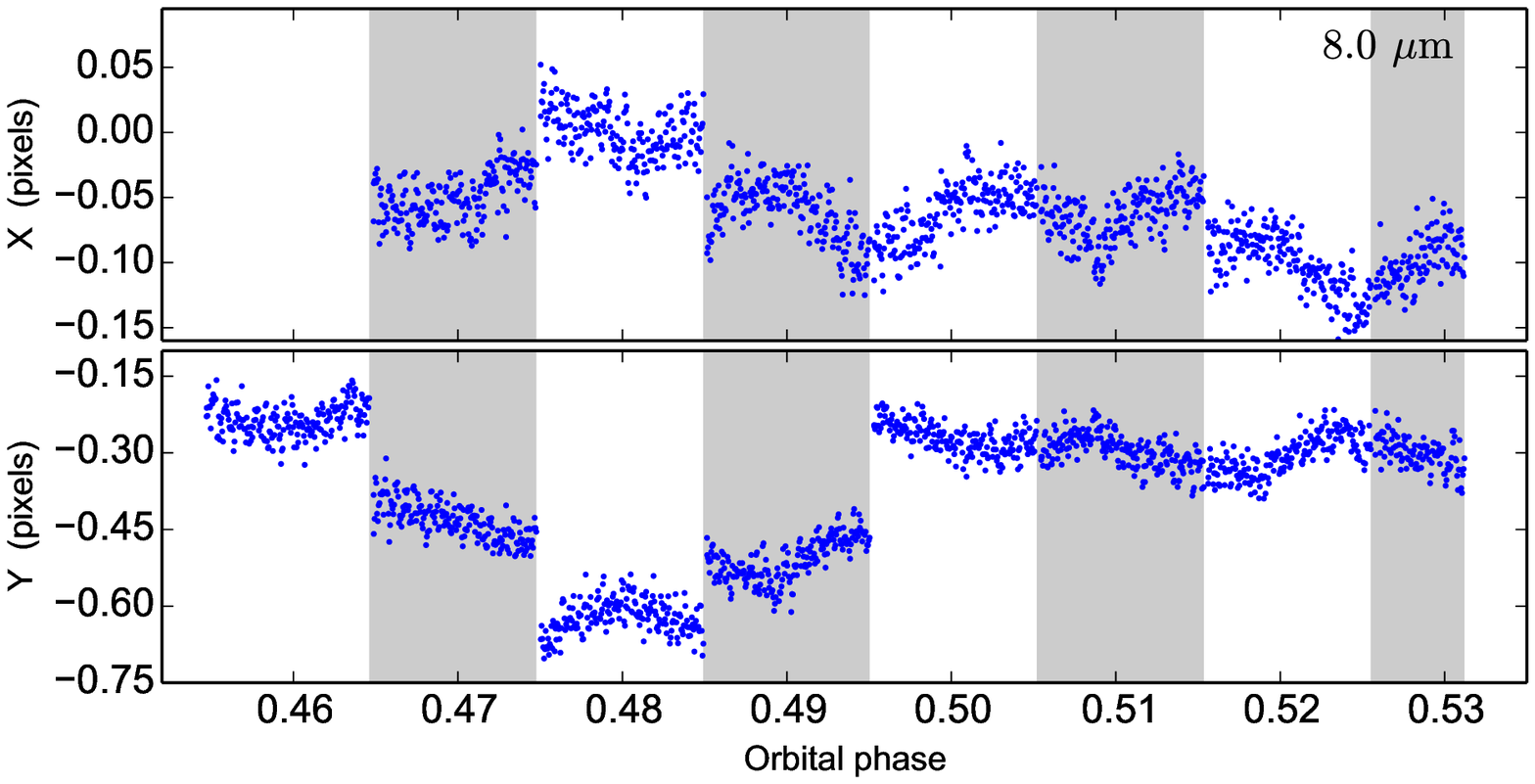}
\caption{{\bf Left:} TrES-1's $x$ (top) and $y$ (bottom) position on
  the detector at 3.6 {\microns} {\vs}\ orbital phase.  The coordinate
  origin denotes the center of the nearest pixel. The shaded/unshaded
  areas mark different AORs.  The ($\sim 0.1$ pixels) pointing offsets
  are clear, as well as the usual hour-long pointing oscillation and
  point-to-point jitter ($\sim 0.01$ pixels). {\bf Right:} Same as the
  left panel, but for the 8.0 {\micron} light curve. The 5.8 and 4.5
  {\micron} datasets were observed simultaneously with the 3.6 and 8.0
  {\micron} bands, respectively; hence, their pointing correlates with
  the ones shown.}
\label{fig:positionAOR}
\end{figure*}

% This work:
This paper analyzes all {\Spitzer} eclipse and transit data for TrES-1
to constrain the planet's orbit, atmospheric thermal profile, and
chemical abundances.  TrES-1's eclipse has an inherently low
signal-to-noise ratio (S/N).  Additionally, as one of the earliest
{\Spitzer} observations, the data did not follow the best observing
practices developed over the years.  We take this opportunity to
present the latest developments in our Photometry for Orbits,
Eclipses, and Transits (POET) pipeline
\citep{StevensonEtal2010natGJ436b, StevensonEtal2012apjHD149026b,
  StevensonEtal2012apjGJ436c, CampoEtal2011apjWASP12b,
  NymeyerEtal2011apjWASP18b, CubillosEtal2013ApjWASP8b} and
demonstrate its robustness on low S/N data.  We have implemented the
differential-evolution Markov-chain Monte Carlo algorithm
\citep[DEMC,][]{Braak2006DifferentialEvolution}, which explores the
parameter phase space more efficiently than the typically-used
Metropolis Random Walk with a multivariate Gaussian distribution as
the proposal distribution.  We also test and compare multiple
centering (Gaussian fit, center of light, PSF fit, and least
asymmetry) and photometry (aperture and optimal) routines.

Section \ref{sec:observations} describes the {\Spitzer} observations.
Section \ref{sec:analysis} outlines our data analysis pipeline.
Section \ref{sec:orbit} presents our orbital analysis.  Section
\ref{sec:atmosphere} shows the constraints that our eclipse
measurements place on TrES-1's atmospheric properties.  Finally,
section \ref{sec:conclusions} states our conclusions.

\section{OBSERVATIONS}
\label{sec:observations}

We analyzed eight light curves of TrES-1 from six {\Spitzer} visits
(obtained during the cryogenic mission): a simultaneous eclipse
observation in the 4.5 and 8.0 {\micron} Infrared Array Camera (IRAC)
bands (PI Charbonneau, program ID 227, full-array mode), a
simultaneous eclipse observation in the 3.6 and 5.8 {\micron} IRAC
bands (PI Charbonneau, program ID 20523, full-array), three
consecutive eclipses in the 16 {\micron} Infrared Spectrograph (IRS)
blue peak-up array, and one transit visit at 16 {\microns} (PI
Harrington, program ID 20605).  Table \ref{table:observations} shows
the {\Spitzer} band, date, total duration, frame exposure time, and
{\Spitzer} pipeline of each observation.

\begin{table}[ht]
\centering
\caption{\label{table:observations} Observation Information}
\strut\hfill
\begin{tabular}{cccccc}
\hline
\hline
Event            & Band    & Observation & Duration & Exp. time & \Spitzer  \\
                 & \microns  & date      &  hours   & seconds   & pipeline  \\
\hline
Eclipse          & \n3.6     & 2005 Sep 17 & 7.27   &  \phn1.2  & S18.18.0  \\ 
Eclipse          & \n4.5     & 2004 Oct 30 & 5.56   &     10.4  & S18.18.0  \\ 
Eclipse          & \n5.8     & 2005 Sep 17 & 7.27   &     10.4  & S18.18.0  \\ 
Eclipse          & \n8.0     & 2004 Oct 30 & 5.56   &     10.4  & S18.18.0  \\ 
Ecl. visit 1     &  16.0     & 2006 May 17 & 5.60   &     31.5  & S18.7.0   \\ 
Ecl. visit 2     &  16.0     & 2006 May 20 & 5.60   &     31.5  & S18.7.0   \\ 
Ecl. visit 3     &  16.0     & 2006 May 23 & 5.60   &     31.5  & S18.7.0   \\ 
Transit          &  16.0     & 2006 May 15 & 5.77   &     31.5  & S18.18.0  \\
\hline
\end{tabular}
\hfill\strut
\end{table}

In 2004, the telescope's Astronomical Observation Request (AOR)
allowed only a maximum of 200 frames
\citep{CharbonneauEtal2005apjTrES1}, dividing the 4.8 and 8.0
{\micron} events into eight AORs (Figure \ref{fig:positionAOR}).  The
later 3.6 and 5.8 {\micron} events consisted of two AORs.  The
repointings between AORs ($\sim 0.1$-pixel offsets) caused systematic
flux variations, because of IRAC's well-known position-dependent
sensitivity variations \citep{CharbonneauEtal2005apjTrES1}.  On the
other hand, the pointing of the IRS observations (a single AOR) cycled
among four nodding positions every five acquisitions, producing flux
variations between the positions.

\section{DATA ANALYSIS}
\label{sec:analysis}

Our POET pipeline processes {Spitzer} Basic Calibrated Data to produce
light curves, modeling the systematics and eclipse (or transit)
signals.  Initially, POET flags bad pixels and calculates the frames'
Barycentric Julian Dates (BJD), reporting the frame mid-times in both
Coordinated Universal Time (UTC) and Barycentric Dynamical Time (TDB).
Next, it estimates the target's center position using any of four
methods: fitting a two-dimensional, elliptical, non-rotating Gaussian
with constant background
\citep[][Supplementary Information]{StevensonEtal2010natGJ436b};
fitting a 100x oversampled point spread function
\citep[PSF,][]{CubillosEtal2013ApjWASP8b}; calculating the center of
light \citep[][]{StevensonEtal2010natGJ436b}; or calculating the least
asymmetry (Lust et al.\ 2014, submitted).  The Gaussian-fit, PSF-fit,
and center-of-light methods considered a 15 pixel square window
centered on the target's peak pixel.  The least-asymmetry method used
a nine pixel square window.

\subsection{Optimal Photometry}
\label{sec:optimal}

POET generates raw light curves either from interpolated aperture
photometry \citep[][sampling a range of aperture radii in 0.25 pixel
  increments]{HarringtonEtal2007natHD149026b} or using an optimal
photometry algorithm (following \citealp{Horne1986Optimal}), which
improves S/N over aperture photometry for low-S/N data sets.  Optimal
photometry has been implemented by others to extract light curves
during stellar occultations by Saturn's rings
\citep{HarringtonEtal2010SatOcculatation} or exoplanets
\citep{Deming2005Nat, StevensonEtal2010natGJ436b}.  This algorithm
uses a PSF model, $P\sb{i}$, to estimate the expected fraction of the
sky-subtracted flux, $F\sb{i}$, falling on each pixel, $i$; divides it
out of $F\sb{i}$ so that each pixel becomes an estimate of the full
flux (with radially increasing uncertainty); and uses a mean with
weights $W\sb{i}$ to give an unbiased estimate of the target flux:
\begin{equation}
f = \frac{\sum\sb{i}W\sb{i}\ F\sb{i}/P\sb{i}}{\sum\sb{i} W\sb{i}}.
\label{eq:opphot1}
\end{equation}
Here, $W\sb{i}=P\sp{2}\sb{i}/V\sb{i}$, with $V\sb{i}$ the variance of
$F\sb{i}$.  Thus,
\begin{equation}
f\sb{\rm opt} = \frac{\sum\sb{i} P\sb{i}\ F\sb{i}/V\sb{i}}
                     {\sum\sb{i} P\sp{2}\sb{i}/V\sb{i}   }.
\label{eq:opphot2}
\end{equation}
We used the Tiny-Tim
program\footnote{\href{http://irsa.ipac.caltech.edu/data/SPITZER/docs/dataanalysistools/tools/contributed/general/stinytim/}{http://irsa.ipac.caltech.edu/data/SPITZER/docs/dataanalysistools/}
  \href{http://irsa.ipac.caltech.edu/data/SPITZER/docs/dataanalysistools/tools/contributed/general/stinytim/}{\hspace{0.3cm}contributed/general/stinytim/}}
(ver. 2.0) to generate a super-sampled PSF model ($100\times$ finer
pixel scale than the {\Spitzer} data).  We shifted the position,
binned down the resolution, and scaled the PSF flux to fit the data.

\subsection{Light Curve Modeling}

Considering the position-dependent (intrapixel) and time-dependent
(ramp) {\Spitzer} systematics \citep{CharbonneauEtal2005apjTrES1}, we
modeled the raw light-curve flux, $F$, as a function of pixel position
$(x,y)$ and time $t$ (in orbital phase units):
\begin{equation}
F(x,y,t) = F\sb{s}\,E(t)\,M(x,y)\,R(t)\,A(a),
\label{eq:lcmodel}
\end{equation}
where $F\sb s$ is the out-of-eclipse system flux (fitting
parameter). $E(t)$ is an eclipse or transit (small-planet
approximation) \citet{MandelAgol2002ApJtransits} model.
$M(x,y)$ is a Bi-Linearly Interpolated Subpixel Sensitivity (BLISS)
map \citep{StevensonEtal2012apjHD149026b}.
$R(t)$ is a ramp model and $A(a)$ a per-AOR flux scaling factor.
The intrapixel effect is believed to originate from non-uniform
quantum efficiency across the pixels
\citep{ReachEtal2005paspIRACcalib}, being more significant at 3.6 and 4.5 {\microns}.  At the longer wavebands, the
intrapixel effect is usually negligible
\citep[e.g.,][]{KnutsonEtal2008apjHD209, KnutsonEtal2011apjGJ436b,
  StevensonEtal2012apjHD149026b}.  The BLISS map outperforms
polynomial fits for removing {\Spitzer}'s position-dependent
sensitivity variations \citep{StevensonEtal2012apjHD149026b,
  BlecicEtal2013apjWASP14b}.

For the ramp systematic, we tested several equations, $R(t)$, from the
literature \citep[e.g.,][]{StevensonEtal2012apjHD149026b,
  CubillosEtal2013ApjWASP8b}.  The data did not support models more
complex than:
\begin{eqnarray}
\label{eq:lin}
{\rm linramp:}  \quad R(t) & = & 1 + r\sb{1}(t-t\sb{c}) \\
\label{eq:quad}
{\rm quadramp:} \quad R(t) & = & 1 + r\sb{1}(t-t\sb{c}) + r\sb{2}(t-t\sb{c})\sp{2} \\
\label{eq:log}
{\rm logramp:}  \quad R(t) & = & 1 + r\sb{1}[\ln(t-t\sb{0})] \\
\label{eq:re}
{\rm risingexp:}\quad R(t) & = & 1 - e\sp{-r\sb{1}(t-t\sb{0})}
\end{eqnarray}
where $t\sb{c}$ is a constant, fixed at orbital phase 0 (for transits)
or 0.5 (for eclipses); $r\sb{1}$ and $r\sb{2}$ are a linear and
quadratic free parameters, respectively; and $t\sb{0}$ is a
time-offset free parameter.

\begin{figure}[tb]
\centering
\includegraphics[width=\linewidth, clip]{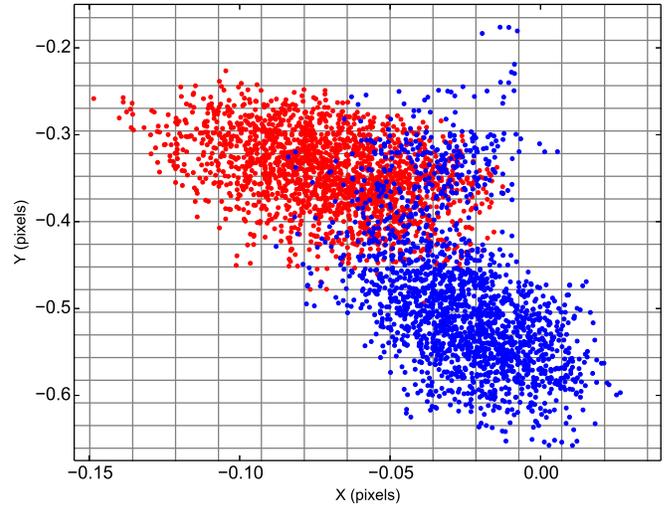}
\caption{3.6 {\micron} detector pointing.  The blue and red points
  denote the data point from the first and second AOR, respectively.
  The coordinate origin denotes the center of the nearest pixel.  The
  grid delimits the BLISS-map bin boundaries.}
\label{fig:BLISSmapch1}
\end{figure}

Additionally, the telescope pointing settled at slightly different
locations for each AOR, resulting in significant non-overlapping regions between
the sets of positions from each AOR (Figure \ref{fig:BLISSmapch1}).
Furthermore, the overlaping region is mostly composed of data points
taken during the telescope settling (when the temporal variation is
stronger).
The pointing offsets provided a weak link between the non-overlapping
regions of the detector, complicating the construction of the pixel
sensitivity map at 3.6 and 4.5 {\microns}.  We attempted the
correction of \citet{StevensonEtal2012apjHD149026b}, $A(a\sb{i})$,
which scales the flux from each AOR, $a\sb{i}$, by a constant factor.
To avoid degeneracy, we set $A(a\sb{1})=1$ and free subsequent
factors.  This can be regarded as a further refinement to the
intrapixel map for 3.6 and 4.5 {\microns}.  Just like the ramp models,
the AOR-scaling model works as an ad-hoc model that corrects for the
{\Spitzer} systematic variations.

Note that introducing
parameters that relate only to a portion of the data violates an
assumption of the Bayesian Information Criterion (BIC) used below; the
same violation occurs for the BLISS map (see Appendix A of
\citealp{StevensonEtal2012apjHD149026b}).  We have not found an
information criterion that handles such parameters, so we ranked these
fits with the others, being aware that BIC penalizes them too harshly.
It turned out that the AOR-scaling model made a significan improvement
only at 3.6~{\microns}; see Section \ref{sec:joint}.

\comment{For this data set, we discarded the other
models by contrasting the estimated orbital parameters to the rest of
the datasets.  At longer wavelengths, the
intrapixel effect is negligible, hence there are no solid grounds to
include the AOR model.}

To determine the best-fitting parameters, ${\bf x}$, of a model, ${\cal M}$
(Equation \ref{eq:lcmodel} in this case), given the data, ${\bf D}$,
we maximize the Bayesian posterior probability \citep[probability of the
model parameters given the data and modeling framework,][]{Gregory2005BayesianBook}:
\begin{equation}
  P({\bf x}|{\bf D},{\cal M}) = P({\bf x}|{\cal M})\, P({\bf D}|{\bf x},{\cal M}) / P({\bf D}|{\cal M}),
\label{eq:bayes}
\end{equation}
where $P({\bf D}|{\bf x},{\cal M})$ is the usual likelihood of the data given the
model and
$P({\bf x}|{\cal M})$ is any prior information on the parameters.  Assuming
Gaussian-distributed priors, maximizing Equation (\ref{eq:bayes}) can
be turned into a problem of minimization:
\begin{equation}
\min \bigg\{
  \sum\sb{j} \left(\frac{{\bf x}\sb{j}   -p\sb{j}}{\sigma\sb{j}}\right)\sp{2} +
  \sum\sb{i} \left(\frac{{\cal M}\sb{i}({\bf x})-{\bf D}\sb{i}}{\sigma\sb{i}}\right)\sp{2} 
     \bigg\},
\label{eq:minimization}
\end{equation}
with $p\sb{j}$ a prior estimation (with standard deviation
$\sigma\sb{j}$).  The second term in Equation (\ref{eq:minimization}) corresponds to $\chi\sp2$.  We used
the Levenberg-Marquardt minimizer to find ${\bf x}\sb{j}$ \citep{Levenberg1944, Marquardt1963}.  Next we sampled the
parameters' posterior distribution through a Markov-chain Monte Carlo
(MCMC) algorithm to estimate the parameter uncertainties, requiring
the Gelman-Rubin statistic \citep{Gelman1992} to be
within 1\% of unity for each free parameter before declaring convergence.

\subsection{Differential Evolution  Markov Chain}
\label{sec:demc}

The MCMC's performance depends crucially on having good proposal
distributions to efficiently explore the parameter space. Previous
POET versions used the Metropolis random walk, where new parameter
sets are proposed from a multivariate normal distribution.
The algorithm's efficiency was limited by the heuristic tuning of the
characteristic jump sizes for each parameter. Too-large values yielded
low acceptance rates, while too-small values wasted computational
power.  Furthermore, highly correlated parameter spaces required
additional orthogonalization techniques
\citep{StevensonEtal2012apjHD149026b} to achieve reasonable acceptance
ratios, and even then did not always converge.

We eliminated the need for manual tuning and orthogonalization
by implementing the differential-evolution Markov-chain algorithm \citep[DEMC,
][]{Braak2006DifferentialEvolution}, which automatically adjusts the
jumps' scales and orientations.  Consider ${\bf x}\sb{n}\sp{i}$ as the set of
free parameters of a chain $i$ at iteration $n$. DEMC runs several
chains in parallel, drawing the parameter values for the next
iteration from the difference between the current parameter states of two
other randomly-selected chains, $j$ and $k$:
\begin{equation}
\label{eq:demc}
{\bf x}\sb{n+1}\sp{i} = {\bf x}\sb{n}\sp{i} + \gamma \left( {\bf x}\sb{n}\sp{j} - {\bf x}\sb{n}\sp{k}\right) + \gamma\sb{2}{\bf e}\sb{n}\sp{i},
\end{equation}
where $\gamma$ is a scaling factor of the proposal jump.  Following
\citet{Braak2006DifferentialEvolution}, we selected $\gamma =
2.38/\sqrt{2d}$ (with $d$ being the number of free parameters) to
optimize the acceptance probability \citep[$\gtrsim $25\%,][]{RobertsEtal1997}.
The last term, $\gamma\sb{2}{\bf e}$,
is a random distribution (of smaller scale than the posterior
distribution) that ensures a complete exploration of posterior
parameter space.  We chose a multivariate normal distribution for {\bf e}, scaled by the factor $\gamma\sb{2}$.

As noted by \citet{EastmanEtal2013paspEXOFAST}, each parameter of $\bf
e$ requires a specific jump scale.  One way to estimate the
scales is to calculate the standard deviation of the parameters in
a sample chain run.
In a second method \citep[similar to that of
][]{EastmanEtal2013paspEXOFAST}, we searched for the limits around the
best-fitting value where $\chi\sp{2}$ increased by 1 along the
parameter axes.  We varied each parameter separately, keeping the
other parameters fixed.  Then, we calculated the jump scale from the
difference between the upper and lower limits, $({\bf x}\sp{\rm
  up}-{\bf x}\sp{\rm lo})/2$.  Both methods yielded similar results in
our tests.
By testing different values for $\gamma\sb{2}$, provided that
$|\gamma\sb{2}{\bf e}\sb{n}\sp{i}| < |\gamma ( {\bf x}\sb{n}\sp{j} -
{\bf x}\sb{n}\sp{k})|$, we found that each trial returned identical
posterior distributions and acceptance rates, so we arbitrarily set
$\gamma\sb{2}=0.1$.

\subsection{Data Set and Model Selection}
\label{sec:modelselec}

To determine the best raw light curve (i.e., the selection of
centering and photometry method), we minimized the standard deviation of
the normalized residuals (SDNR) of the light-curve fit
\citep{CampoEtal2011apjWASP12b}.  This naturally prefers good fits and
low-dispersion data.

We use Bayesian hypothesis testing to select the model best supported
by the data. Following \citet{Raftery1995BIC}, when comparing two
models ${\cal M}\sb{1}$ and ${\cal M}\sb{2}$ on a data set ${\bf D}$, the posterior odds
($B\sb{21}$, also known as Bayes factor) indicates the model preferred
by the data and the extent to which it is preferred.  Assuming that
either model is, a priori, equally probable, the posterior odds are
given by:
\begin{equation}
\label{eq:postodds}
 B\sb{21} =       \frac{p({\bf D}|{\cal M}\sb{2})}{p({\bf D}|{\cal M}\sb{1})}
          =       \frac{p({\cal M}\sb{2}|{\bf D})}{p({\cal M}\sb{1}|{\bf D})}
          \approx \exp\left(-\frac{{\rm BIC}\sb{2}-{\rm BIC}\sb{1}}{2}\right).
\end{equation}
This is the ${\cal M}\sb{2}$-to-${\cal M}\sb{1}$ probability ratio for
the models (given the data), with BIC = $\chi\sp{2} + k\ln{N}$ the
Bayesian Information Criterion \citep{Liddle2007mnrasBIC}, $k$ the
number of free parameters, and $N$ the number of points.  Hence,
${\cal M}\sb{2}$ has a fractional probability of
\begin{equation}
\label{eq:fracprob}
p({\cal M}\sb{2}|{\bf D}) = \frac{1}{1+ 1/B\sb{21}}.
\end{equation}
We selected the best models as those with the lowest BIC, and assessed
the fractional probability of the others (with respect to the best
one) using Equation (\ref{eq:fracprob}).

Recently, \citet{Gibson2014mnrasInferenceSystematics} proposed to
marginalize over systematics models rather than use model selection.
Although this process is still subjected to the researcher's choice of
systematics models to test, it is a more robust method.
Unfortunately, unless we understand the true nature of the systematics
to provide a physically motivated model, the modeling process will
continue to be an arbitrary procedure.  Most of our analyses prefer
one of the models over the others.  When a second model shows a
significant fractional probability ($>0.2$) we reinforce our selection
based on additional evidence (is the model physically plausible? or
how do the competing models perform in a joint fit?).  We are
evaluating to include the methods of
\citet{Gibson2014mnrasInferenceSystematics} to our pipeline in the
future.

\subsection{Light-curve  Analyses} % :::::::::::::::::::::

We initially fit the eclipse light curves individually to determine
the best data sets (centering and photometry methods) and systematics
models.  Then, we determined the definitive parameters from a final
joint fit (Section \ref{sec:joint}) with shared eclipse parameters.
For the eclipse model we fit the midpoint, depth, duration, and
ingress time (while keeping the egress time equal to the ingress
time).  Given the low S/N of the data, the individual events do not
constrain all the eclipse parameters well.  However, the final joint
fit includes enough data to do the job.  For the individual fits, we
assumed a negligible orbital eccentricity, as indicated by transit and
radial-velocity (RV) data, and used the transit duration ($2.497 \pm
0.012$ hr) and transit ingress/egress time ($18.51 \pm 0.63$ min) from
\citet{WinnEtal2007apjTres1} as priors on the eclipse duration and
ingress/egress time.  In the final joint-fit experiments, we freed
these parameters.

\subsubsection{IRAC - 3.6 $\mu$m Eclipse} % :::::::::::::::::::::
\label{sec:tr001bs11}

\begin{figure}[tb]
\centering
\includegraphics[width=\linewidth, clip]{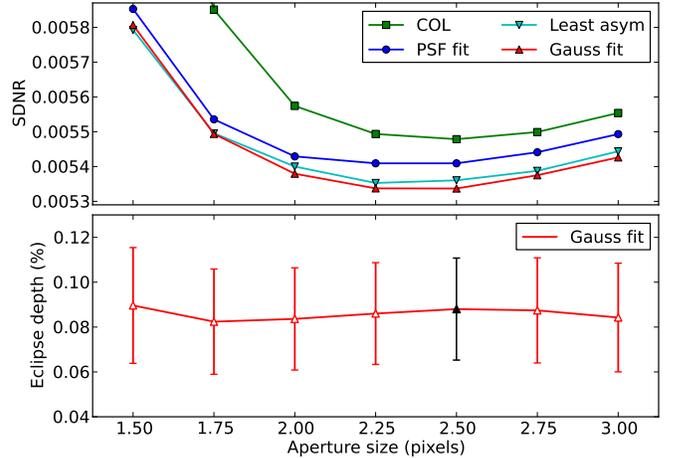}
\caption{{\bf Top:} 3.6 {\micron} eclipse light-curve SDNR {\vs}\ 
  aperture.  The legend indicates the centering method used.  All curves used the best ramp model from Table
  \ref{table:tr001bs11ramps}. {\bf Bottom:} Eclipse depth {\vs}\ 
  aperture for Gaussian-fit centering, with the best aperture (2.5 pixels)
  in black.}
\label{fig:sdnr11}
\end{figure}

This observation is divided into two AORs at phase 0.498, causing a
systematic flux offset due to IRAC's intrapixel sensitivity
variations.
We tested aperture photometry between 1.5 and 3.0 pixels.
The eclipse depth is consistent among the apertures, and the
minimum SDNR occurs for the 2.5 pixel aperture with Gaussian-fit
centering (Figure \ref{fig:sdnr11}).

Table \ref{table:tr001bs11ramps} shows the best four model fits at the
best aperture; $\Delta$BIC is the BIC difference with respect to the
lowest BIC.  Given the relatively large uncertainties, more-complex
models are not supported, due to the penalty of the additional free
parameters.  The Bayesian Information Criterion favors the AOR-scaling
model (Table \ref{fig:sdnr11}, last column).

Although the fractional probabilities of the quadratic and exponential
ramp models are not negligible, we discard them based on the estimated
midpoints, which differ from a circular orbit by 0.008 (twice the
ingress/egress duration).
It is possible that a non-uniform brightness distribution can induce
offsets in the eclipse midpoint \citep{WilliamsEtal2006apj}, and these
offsets can be wavelength dependent.  However, this relative offset can be
at most the duration of the ingress/egress.
Therefore, disregarding non-uniform brightness offsets, considering
the lack of evidence for transit-timing variations and that all other
data predict a midpoint consistent with a circular orbit, the
3.6~{\micron} offset must be caused by systematic effects.
The AOR-scaling model is the only one that yields a midpoint
consistent with the rest of the data.  Our joint-fit analysis (Section
\ref{sec:joint}) will provide further support to our model selection.

\renewcommand{\tabcolsep}{4pt}
\begin{table}[ht]
\caption{3.6-{\microns} eclipse - ramp model fits\protect\footnotemark}
\label{table:tr001bs11ramps}
\strut\hfill\begin{tabular}{cccccc}
\hline
\hline
$R(t)\,A({\rm a})$ & Ecl. Depth & Midpoint & SDNR  & $\Delta$BIC & $p({\cal M}\sb{2}|D)$ \\
          & (\%)\footnotemark   & (phase)  &       &             &      \\
\hline
$A(a)$    & 0.083(24)  & 0.501(4)   & 0.0053763 &   0.0  & $\cdots$     \\
quadramp  & 0.158(29)  & 0.492(2)   & 0.0053712 &   2.8  & 0.19         \\
risingexp & 0.146(25)  & 0.492(2)   & 0.0053715 &   2.9  & 0.19         \\
linramp   & 0.093(23)  & 0.492(3)   & 0.0053814 &   7.4  & 0.02         \\
\hline
\end{tabular}\hfill\strut
\footnotetext[1]{Fits for Gaussian-fit centering and 2.5-pixel aperture photometry.}
\footnotetext[2]{For this and the following tables, the values quoted in parenthesis indicate the 1$\sigma$ uncertainty corresponding to the least significant digits.}
\end{table}
\renewcommand{\tabcolsep}{5pt}

\begin{figure*}[p]
\strut\hfill % IRAC raw
\includegraphics[width=0.32\textwidth, clip]{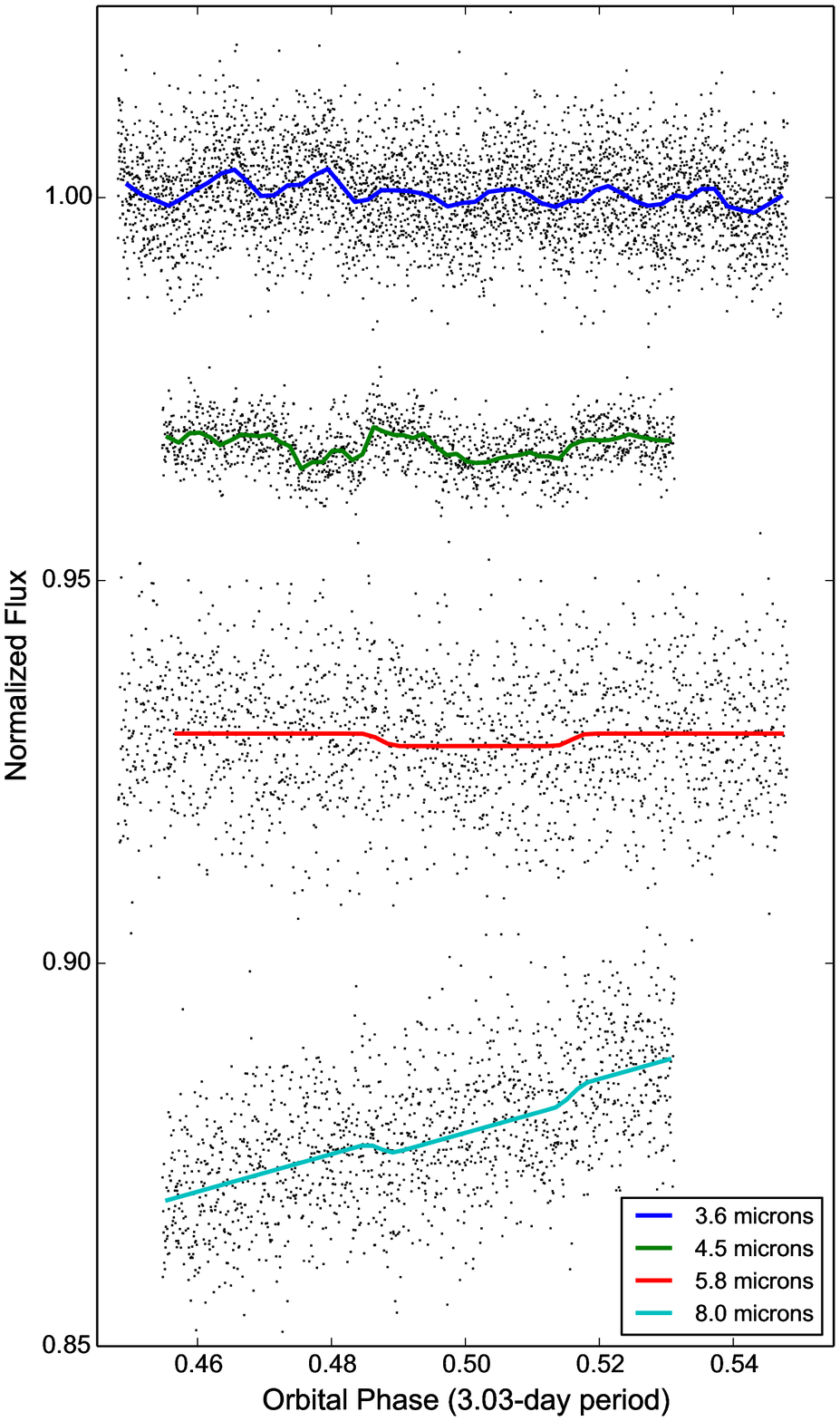}
\hfill       % IRS  raw
\includegraphics[width=0.32\textwidth, clip]{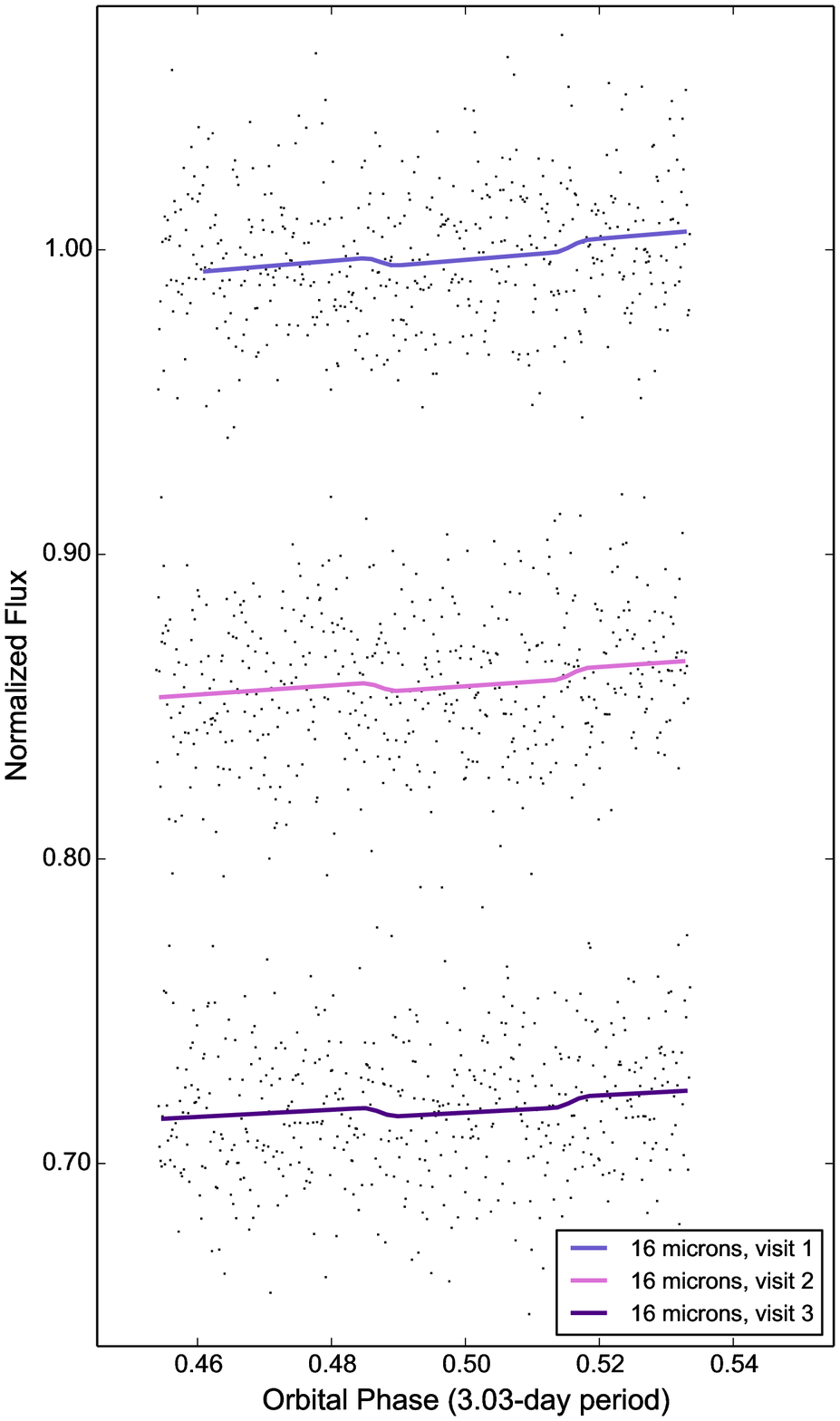}
\hfill       % IRAC bin
\includegraphics[width=0.32\textwidth, clip]{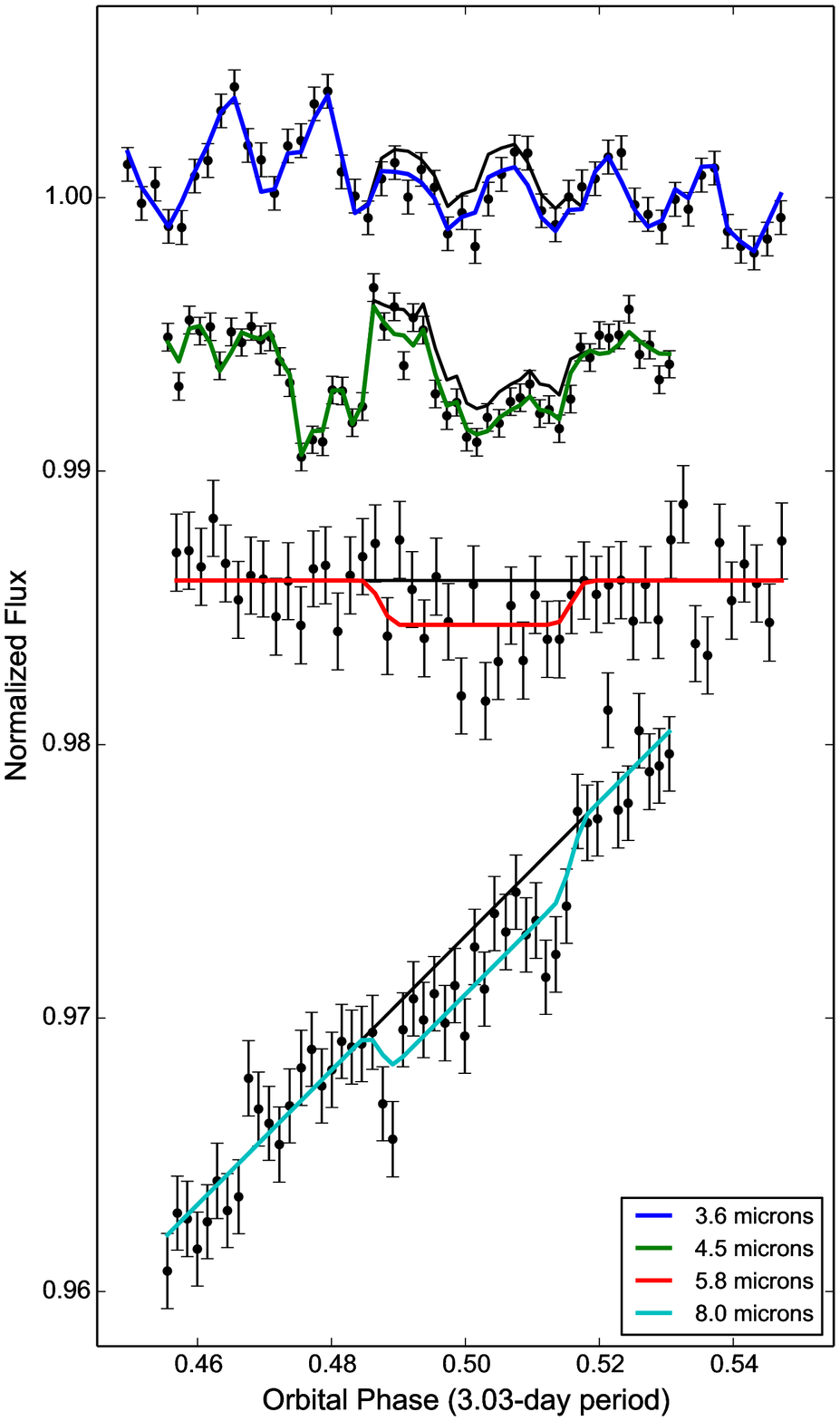}
\hfill\strut
\newline
\strut\hfill % All norm
\includegraphics[width=0.32\textwidth, clip]{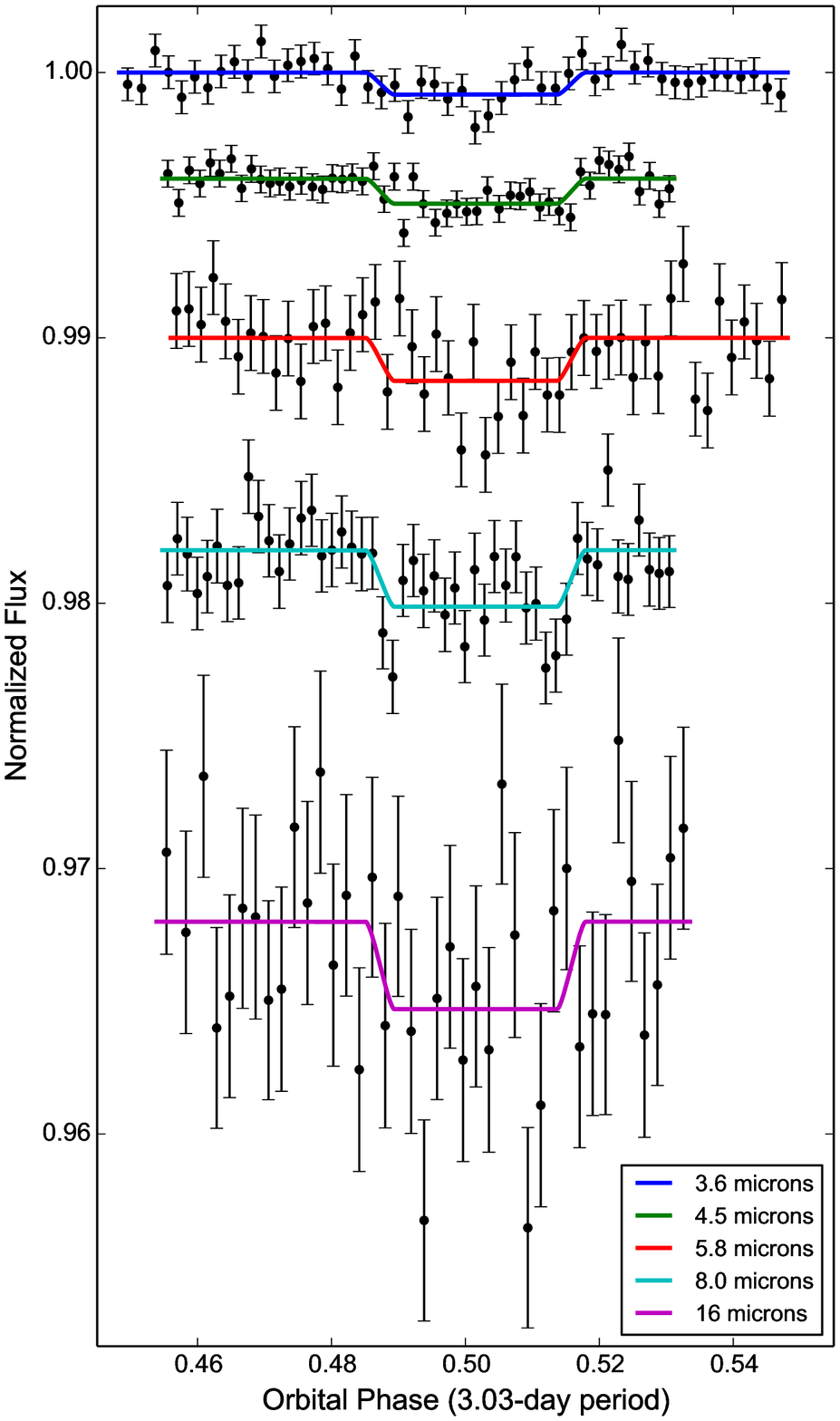}
\hfill       % IRAC RMS
\includegraphics[width=0.32\textwidth, clip]{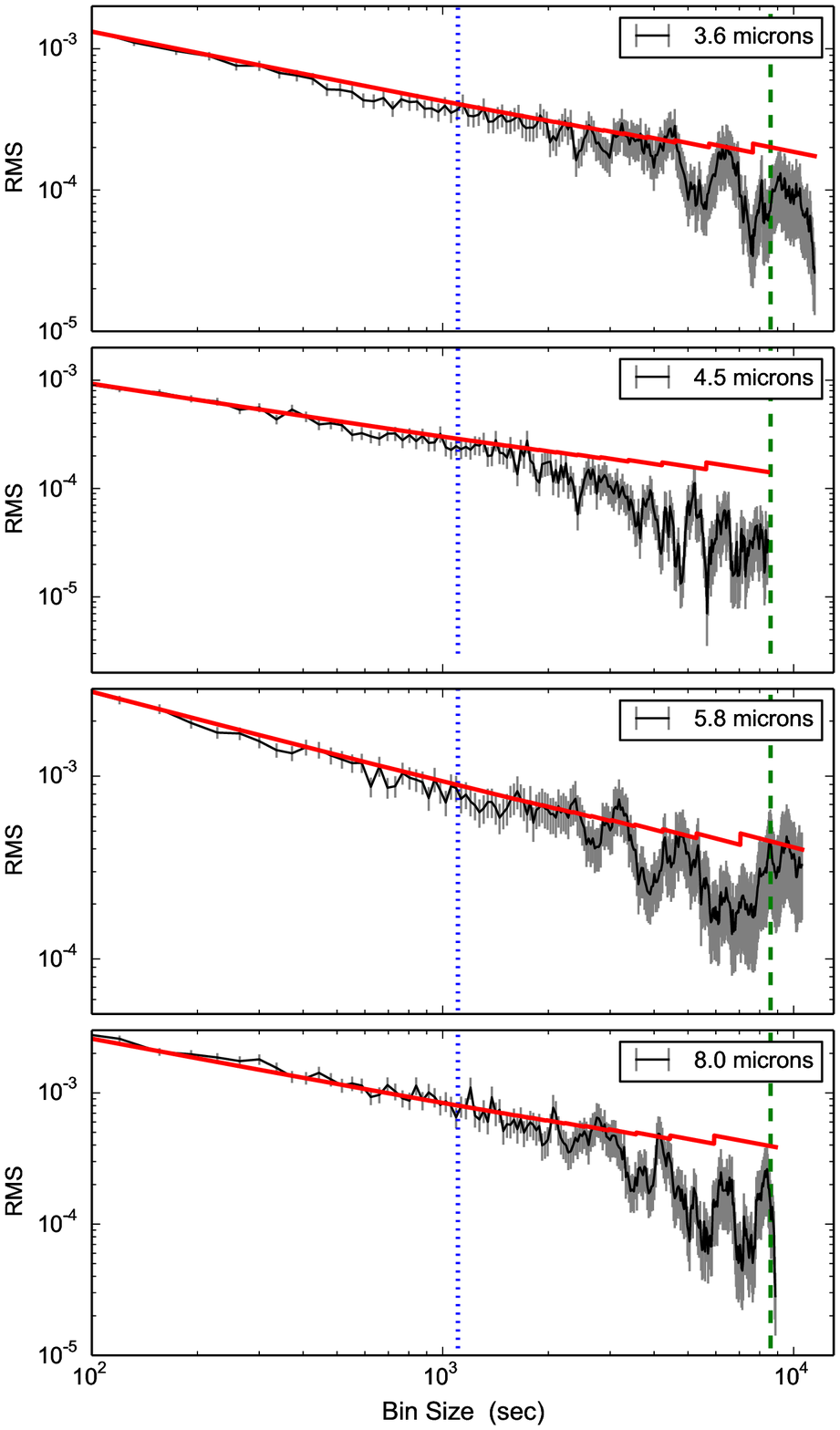}
\hfill       % IRS RMS
\includegraphics[width=0.32\textwidth, clip]{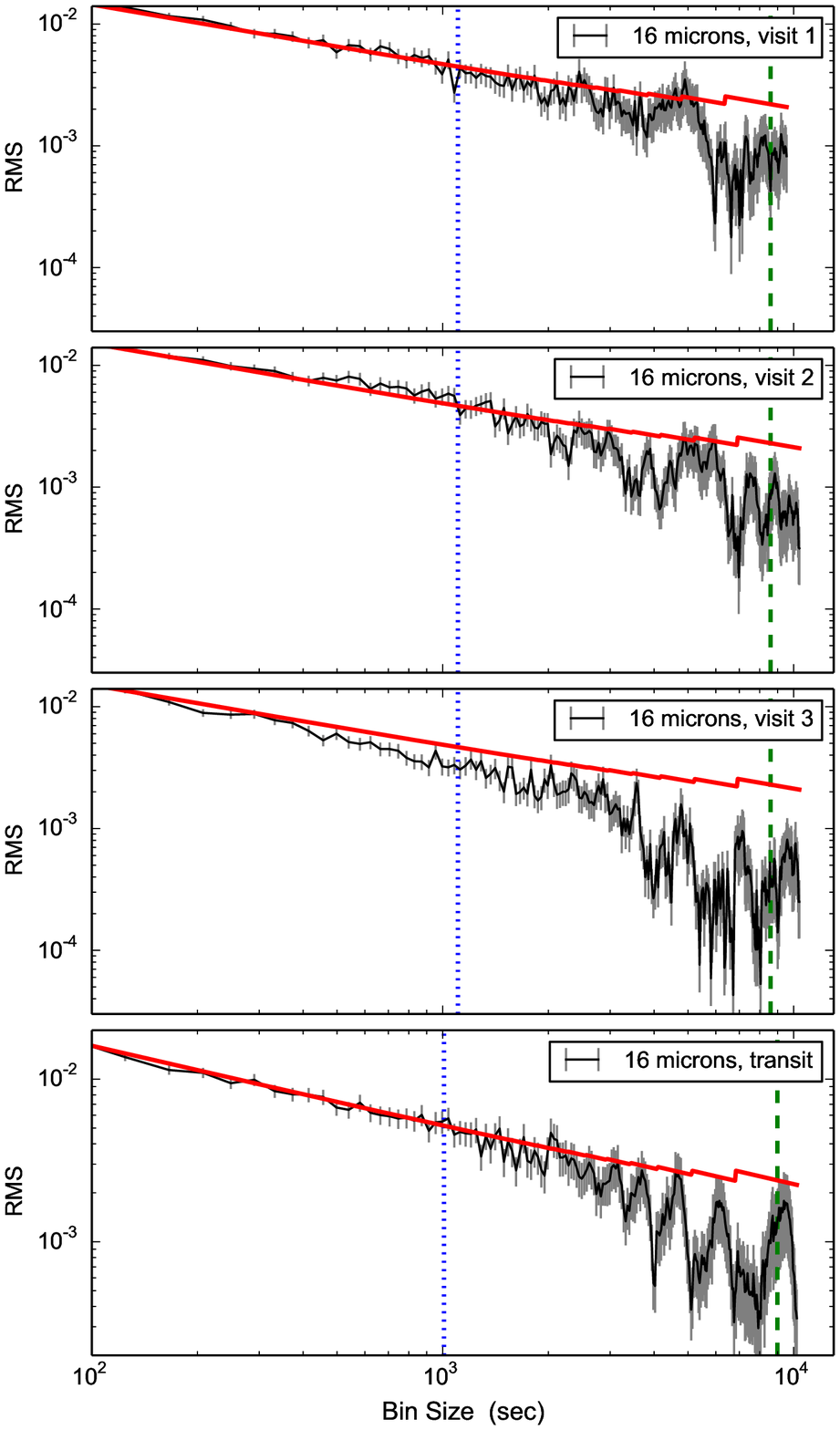}
\hfill\strut
\caption{TrES-1 secondary-eclipse light curves and rms-{\vs}-bin size
  plots.  Raw light curves are in the top-left and top-center panels.
  Binned IRAC data are in the top-right panel, and
  systematics-corrected traces are in the bottom-left panel.  The
  system flux is normalized and the curves are shifted vertically for
  clarity.  The colored solid curves are the best-fit models, while
  the black solid curves are the best-fit models excluding the eclipse
  component.  The error bars give the 1$\sigma$ uncertainties.  The
  bottom-center and bottom-right panels show the fit residuals' rms
  (black curves with 1$\sigma$ uncertainties) {\vs} bin size.  The red
  curves are the expected rms for Gaussian noise.  The blue dotted and
  green dashed vertical lines mark the ingress/egress time and eclipse
  duration, respectively.}
\label{fig:lightcurves}
\end{figure*}

We adjusted the BLISS map model following
\citet{StevensonEtal2012apjHD149026b}.  For a minimum of 4 points per
bin, the eclipse depth remained constant for BLISS bin sizes similar
to the rms of the frame-to-frame position difference (0.014 and 0.026
pixels in $x$ and $y$, respectively).  
Figure \ref{fig:lightcurves}
shows the raw, binned, and systematics-corrected light curves with
their best-fitting models.  

To estimate the contribution from time-correlated residuals we
calculated the time-averaging rms-vs.-bin-size curves
\citep{PontEtal2006mnrasRednoise, WinnEtal2008apjRednoise}.  This
method compares the binned-residuals rms to the uncorrelated-noise
(Gaussian noise) rms.  An excess rms over the Gaussian rms would
indicate a significant contribution from time-correlated residuals.
Figure \ref{fig:lightcurves} (bottom-center and bottom-right panels)
indicates that time-correlated noise is not significant at any
time scale, for any of our fits.

\subsubsection{IRAC - 4.5 $\mu$m Eclipse} % ::::::::::::::::::::::::::
\label{sec:tr001bs21}

\begin{figure}[tb]
\centering
\includegraphics[width=0.85\linewidth, clip]{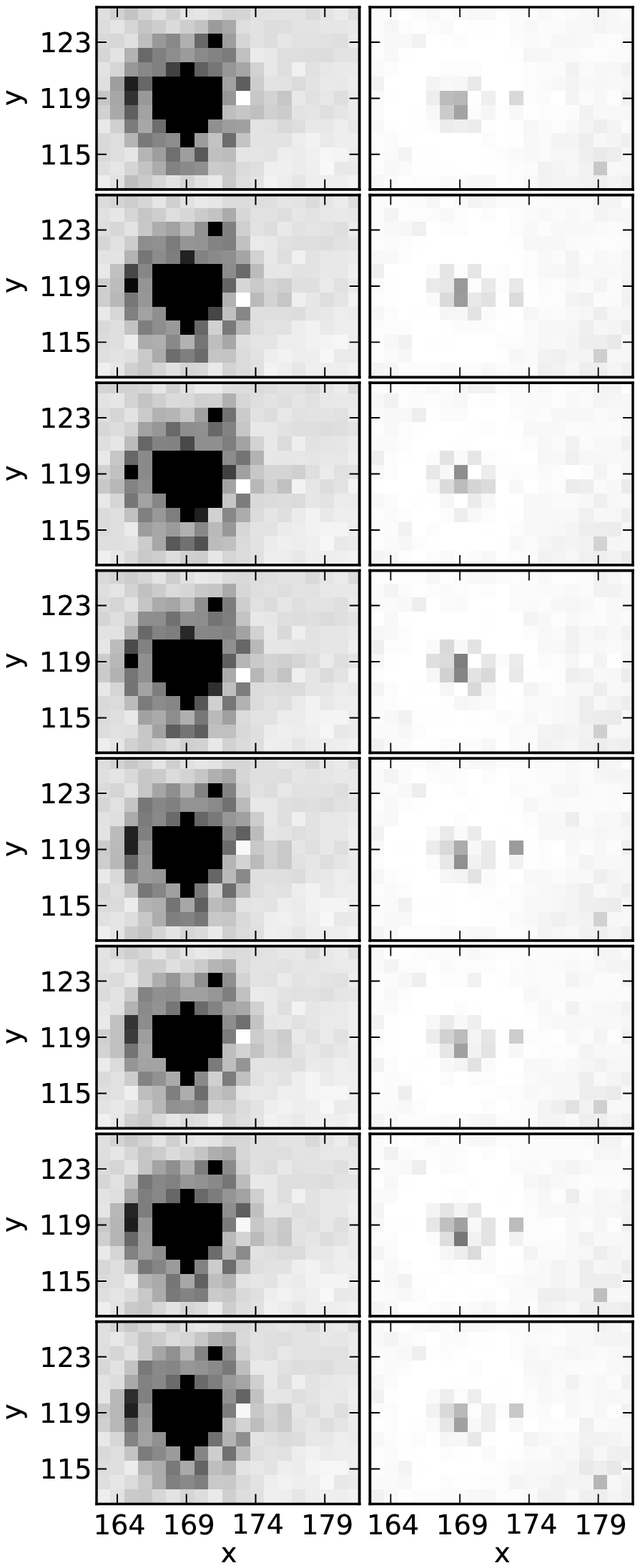}
\caption{ {\bf Left:} Per-AOR mean of the {\Spitzer} BCD frames at 4.5
  {\microns} around TrES-1.  {\bf Right:} Per-AOR rms divided by the
  square root of the mean BCD frames at 4.5 {\microns}.  The flux is
  in electron counts, the color scales range from the 2.5th (white) to
  the 97.5th (black) percentile of the flux distribution.  TrES-1's
  center is located near $x=169, y=119$.  The miscalculated
  muxbleed-corrected pixels stand out at 4 and (sometimes) 8 pixels to
  the right of the target center.  The excess in the scaled rms
  confirms that the muxbleed correction is not linearly scaled with
  the flux.  The pixels around the target center also show high rms
  values, which might be due to the 0.2 pix motion of the PSF
  centers.}
\label{fig:bleed}
\end{figure}

Our analysis of the archival data revealed that the 4.5 {\micron} data
suffered from multiplexer bleed, or ``muxbleed'', indicated by flagged
pixels near the target in the mask frames and data-frame headers
indicating a muxbleed correction.
Muxbleed is an effect observed in the IRAC InSb arrays (3.6 and 4.5
{\microns}) wherein a bright star trails in the fast-read direction
for a large number of consecutive readouts.  Since there are 4 readout
channels, the trail appears every 4 pixels, induced by one or more bright pixels\footnote{
\href{http://irsa.ipac.caltech.edu/data/SPITZER/docs/irac/iracinstrumenthandbook/59/}{http://irsa.ipac.caltech.edu/data/SPITZER/docs/irac/} \\
\href{http://irsa.ipac.caltech.edu/data/SPITZER/docs/irac/iracinstrumenthandbook/59/}{\hspace{0.3cm}iracinstrumenthandbook/59/}} (Figure \ref{fig:bleed}).

TrES-1 (whose flux was slightly below the nominal saturation limit at 4.5
{\microns}) and a second star that is similarly bright fit the muxbleed
description.  We noted the same feature in the BCD frames used by
\citet[][{\Spitzer} pipeline version
S10.5.0]{CharbonneauEtal2005apjTrES1}.  Their headers indicated a
muxbleed correction as well, but did not clarify whether or not a
pixel was corrected.
\comment{From communications at that time, the PI and the IRAC team
concluded that the flagged data were not invalid.}

Since the signal is about $\ttt{-3}$ times the stellar flux level,
every pixel in the aperture is significant and any imperfectly made
local correction raises concern (this is why we do not interpolate bad
pixels in the aperture, but rather discard frames that have them).
Nevertheless, we analyzed the data, ignoring the muxbleed flags, to
compare it to the results of \citet{CharbonneauEtal2005apjTrES1}.  In
the atmospheric analysis that follows, we model the planet both with
and without this data set.

\begin{figure}[ht]
\centering
\includegraphics[width=\linewidth, clip]{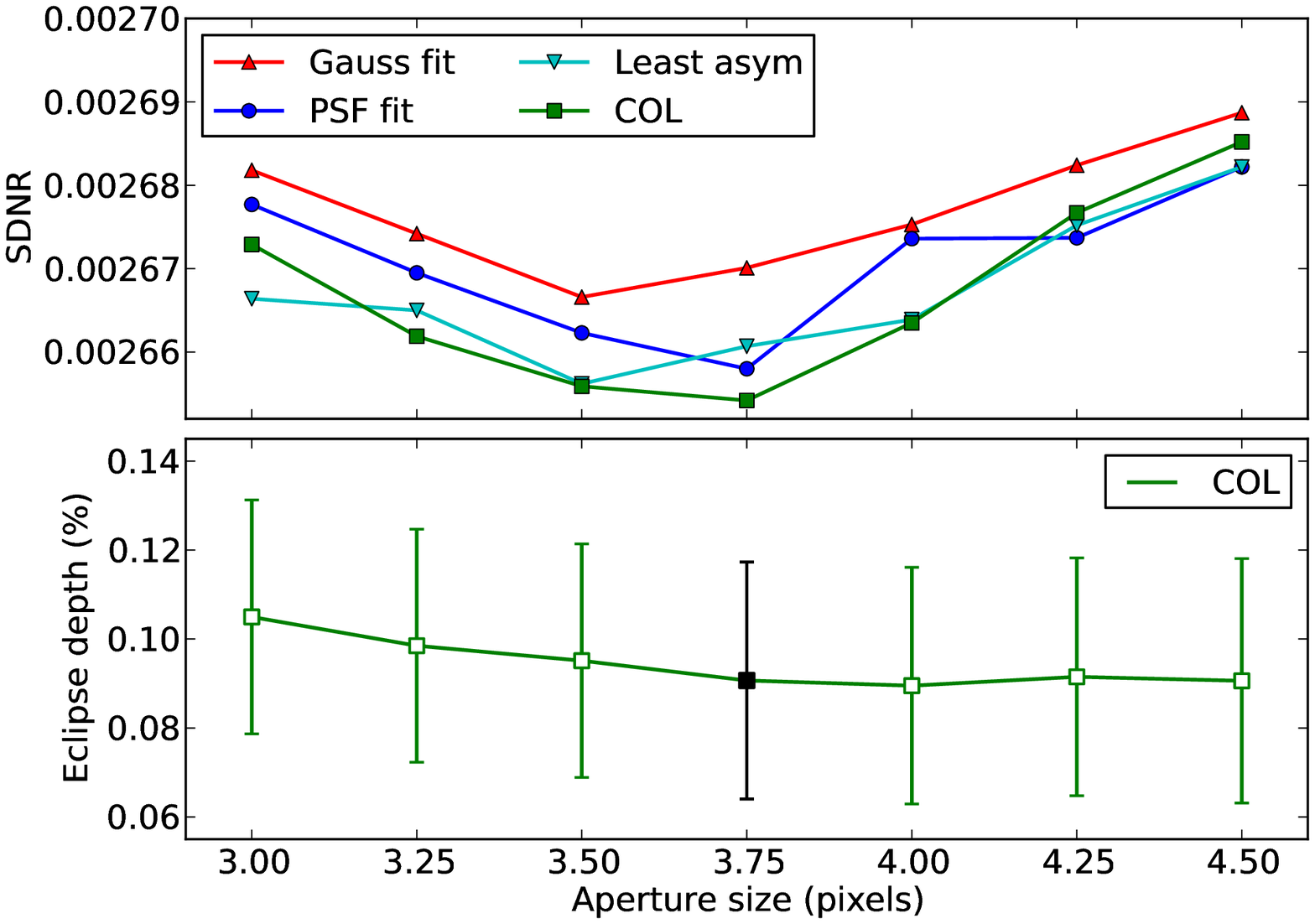}
\caption{{\bf Top:} 4.5 {\micron} eclipse light-curve SDNR {\vs} aperture.
  The legend indicates the centering method used.  All curves used the best ramp model from Table
  \ref{table:tr001bs21ramps}. {\bf Bottom:} Eclipse depth {\vs} aperture
  for center-of-light centering, with the best aperture (3.75 pixels) in black.}
\label{fig:sdnrdepth21}
\end{figure}

This light curve is also mainly affected by the intrapixel effect.
Since the 4.5~{\microns} light curve consisted of 8 AORs, some of
which are entirely in- or out-of-eclipse, making the AOR-scale model
to overfit the data. We tested apertures
between 2.5 and 4.5 pixels, finding the lowest SDNR for the
center-of-light centering method at the 3.75-pixel aperture (Figure
\ref{fig:sdnrdepth21}).  This alone is surprising, as it may be the
first time in our experience that center of light is the best method.
In the same manner as for the 3.6~{\micron} data, we selected BLISS
bin sizes of 0.018 ($x$) and 0.025 ($y$) pixels, for 4 minimum points
per bin.  A fit with no ramp model minimized BIC (Table
\ref{table:tr001bs21ramps}). Figure \ref{fig:lightcurves} shows the
data and best-fitting light curves and the rms-{\vs}-bin size plot.

\begin{table}[ht]
\caption{4.5-{\micron} Eclipse  - ramp model fits\protect\footnotemark}
\label{table:tr001bs21ramps}
\strut\hfill\begin{tabular}{cccccc}
\hline
\hline
$R(t)\,A({\rm a})$  & Ecl. Depth (\%) & SDNR  & $\Delta$BIC & $p({\cal M}\sb{2}|D)$ \\
\hline
no-model            & 0.090(28)       & 0.0026543 & \n0.0   & $\cdots$     \\
linramp             & 0.091(27)       & 0.0026531 & \n6.0   & 0.05         \\
risingexp           & 0.131(32)       & 0.0026469 & \n7.1   & 0.03         \\
quadamp             & 0.153(39)       & 0.0026481 & \n8.7   & 0.01         \\
logramp             & 0.090(22)       & 0.0026532 &  13.3   & $1\tttt{-3}$ \\
$A(a)$              & 0.140(43)       & 0.0026474 &  38.5   & $4\tttt{-9}$ \\
\hline
\end{tabular}\hfill\strut
\footnotetext[1]{Fits for center-of-light centering and 3.75-pixel aperture photometry.}
\end{table}

\subsubsection{IRAC - 5.8 $\mu$m Eclipse} % ::::::::::::::::::::::::::::
\label{sec:tr001bs31}

\begin{figure}[ht]
\centering
\includegraphics[width=\linewidth, clip]{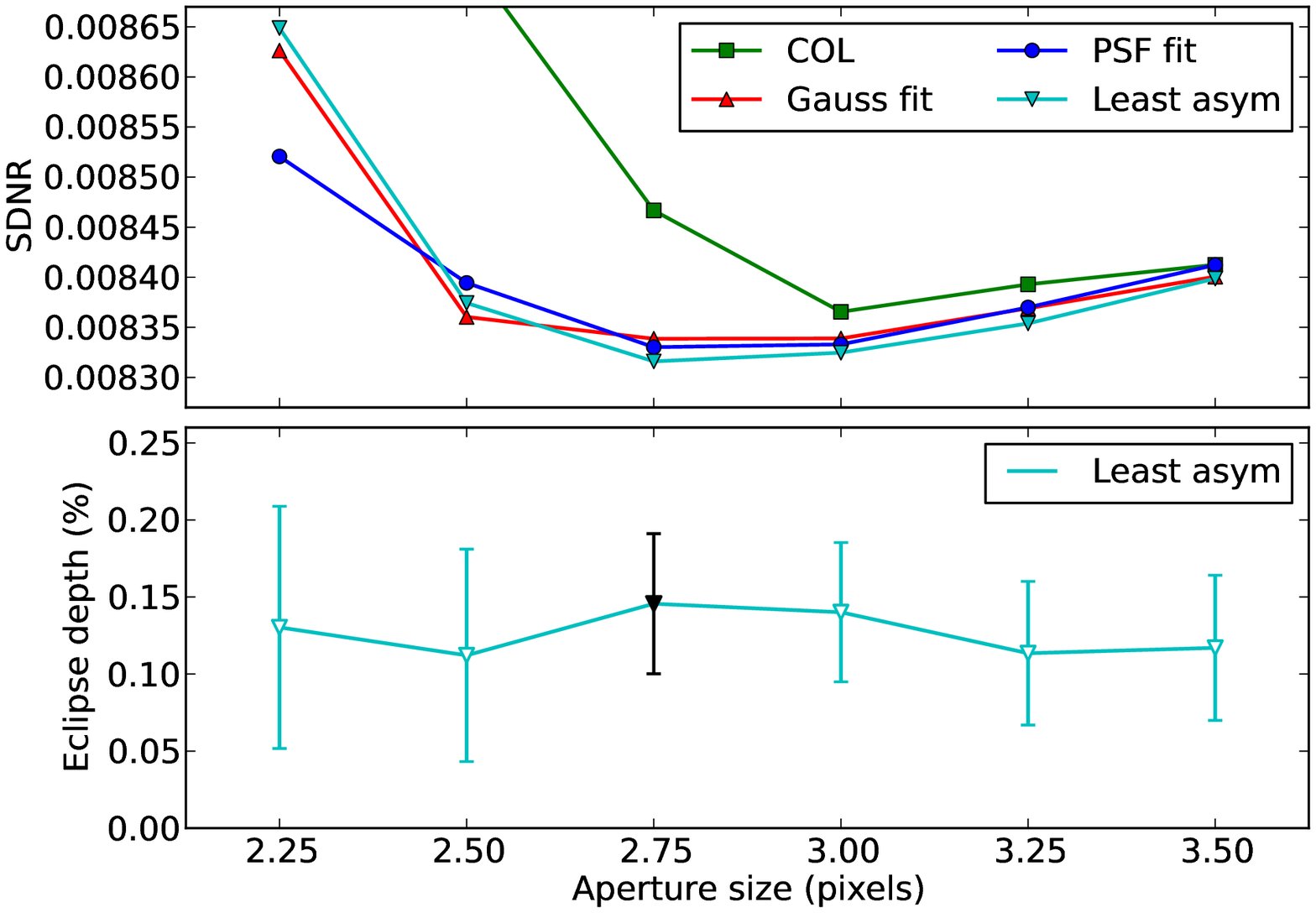}
\caption{
{\bf Top:} 5.8 {\micron} eclipse light-curve SDNR {\vs} aperture.
  The legend indicates the centering method used.  All curves used the best ramp model from Table
  \ref{table:tr001bs31ramps}. {\bf Bottom:} Eclipse depth {\vs} aperture
  for least-asymmetry centering, with the best aperture (2.75 pixels) in black.}
\label{fig:sdnrdepth31}
\end{figure}

These data are not affected by the intrapixel effect.  We sampled
apertures between 2.25 and 3.5 pixels.  Least-asymmetry centering
minimized the SDNR at 2.75 pixels,
with all apertures returning consistent eclipse depths (Figure
\ref{fig:sdnrdepth31}).
The BIC comparison favors a fit without AOR-scale nor ramp models,
although, at some apertures the midpoint posterior distributions
showed a hint of bi-modality.  The eclipse depth, however, remained
consistent for all tested models (Table \ref{table:tr001bs31ramps}).
Figure \ref{fig:lightcurves} shows the data and best-fitting light
curves and rms-{\vs}-bin size plot.

\begin{table}[ht]
\caption{5.8-{\micron} Eclipse - ramp model fits\protect\footnotemark}
\label{table:tr001bs31ramps}
\strut\hfill\begin{tabular}{cccccc}
\hline
\hline
$R(t)\,A({a})$  & Ecl. Depth (\%) & SDNR       & $\Delta$BIC & $p({\cal M}\sb{2}|D)$ \\
\hline
no-model        &   0.158(44)     &  0.0083287 &  \n0.0    & $\cdots$     \\
$A(a)$          &   0.142(45)     &  0.0083220 &  \n4.4    & 0.10         \\
linramp         &   0.154(44)     &  0.0083281 &  \n7.2    & 0.03         \\
quadramp        &   0.100(54)     &  0.0083259 &   13.0    & $2\tttt{-3}$ \\
risingexp       &   0.158(44)     &  0.0083287 &   14.9    & $6\tttt{-4}$ \\
\hline
\end{tabular}\hfill\strut
\footnotetext[1]{Fits for least-asymmetry centering and 2.75-pixel aperture photometry.}
\end{table}

\subsubsection{IRAC - 8.0 $\mu$m Eclipse} % :::::::::::::::::::::
\label{sec:tr001bs41}

\begin{figure}[th]
\centering
\includegraphics[width=\linewidth, clip]{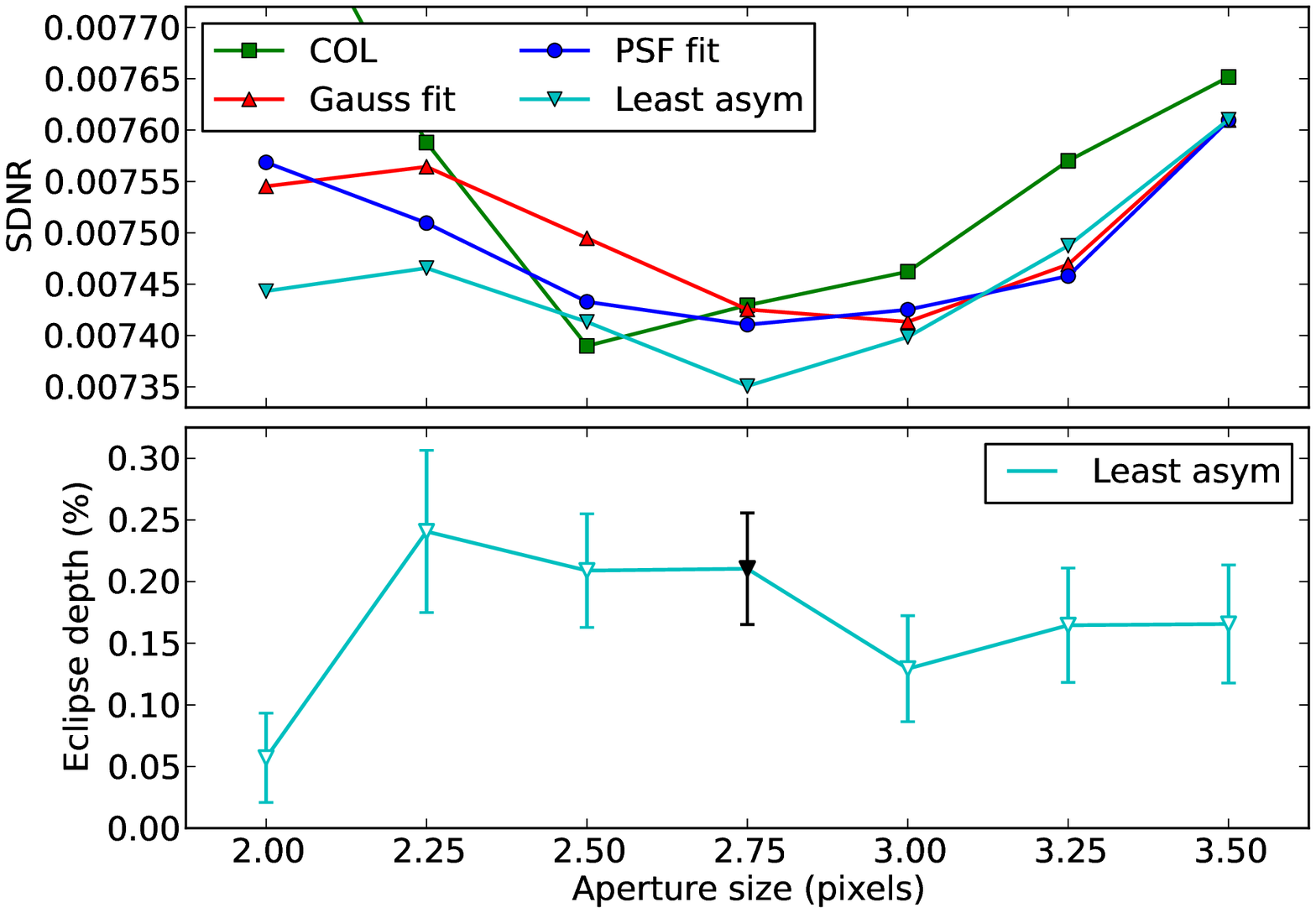}
\caption{{\bf Top:} 8.0 {\micron} eclipse light-curve SDNR {\vs} aperture.
  The legend indicates the centering method used.  All curves used the
  best ramp model from Table
  \ref{table:tr001bs41ramps}. {\bf Bottom:} Eclipse depth {\vs}
  aperture for least-asymmetry centering, with the best aperture (2.75
  pixels) in black.  This data set had the greatest eclipse-depth
  variations per aperture.}
\label{fig:sdnrdepth41}
\end{figure}

This data set had eight AOR blocks.  We tested aperture photometry
from 1.75 to 3.5 pixels.  Again, least-asymmetry centering minimized
the SDNR for the 2.75-pixel aperture (Figure \ref{fig:sdnrdepth41}).
We attempted fitting with the per-AOR adjustment $A(a)$, but the seven
additional free parameters introduced a large BIC penalty, and the
many parameters certainly alias with the eclipse.  The linear ramp
provided the lowest BIC (Table \ref{table:tr001bs41ramps}).  Figure
\ref{fig:lightcurves} shows the data and best-fitting light curves and
rms-{\vs}-bin size plot.

\begin{table}[ht]
\caption{8.0-{\microns} Eclipse - ramp model fits\protect\footnotemark}
\label{table:tr001bs41ramps}
\strut\hfill\begin{tabular}{cccccc}
\hline
\hline
$R(t)\,A({a})$  & Ecl. Depth (\%) & SDNR     & $\Delta$BIC & $p({\cal M}\sb{2}|D)$ \\
\hline
linramp         & 0.208(45)      & 0.0073506 & \n0.0       & $\cdots$      \\
quadramp        & 0.267(62)      & 0.0073388 & \n3.3       & 0.16          \\
risingexp       & 0.278(53)      & 0.0073389 & \n3.4       & 0.15          \\
logramp         & 0.304(45)      & 0.0073471 & \n7.0       & 0.03          \\
linramp--$A(a)$ & 0.759(185)     & 0.0073112 &  41.8       & $8\tttt{-10}$ \\
\hline
\end{tabular}\hfill\strut
\footnotetext[1]{Fits for least-asymmetry centering and 2.75-pixel aperture photometry.}
\end{table}

\begin{figure*}[thb]
\centering
\includegraphics[width=\linewidth, clip]{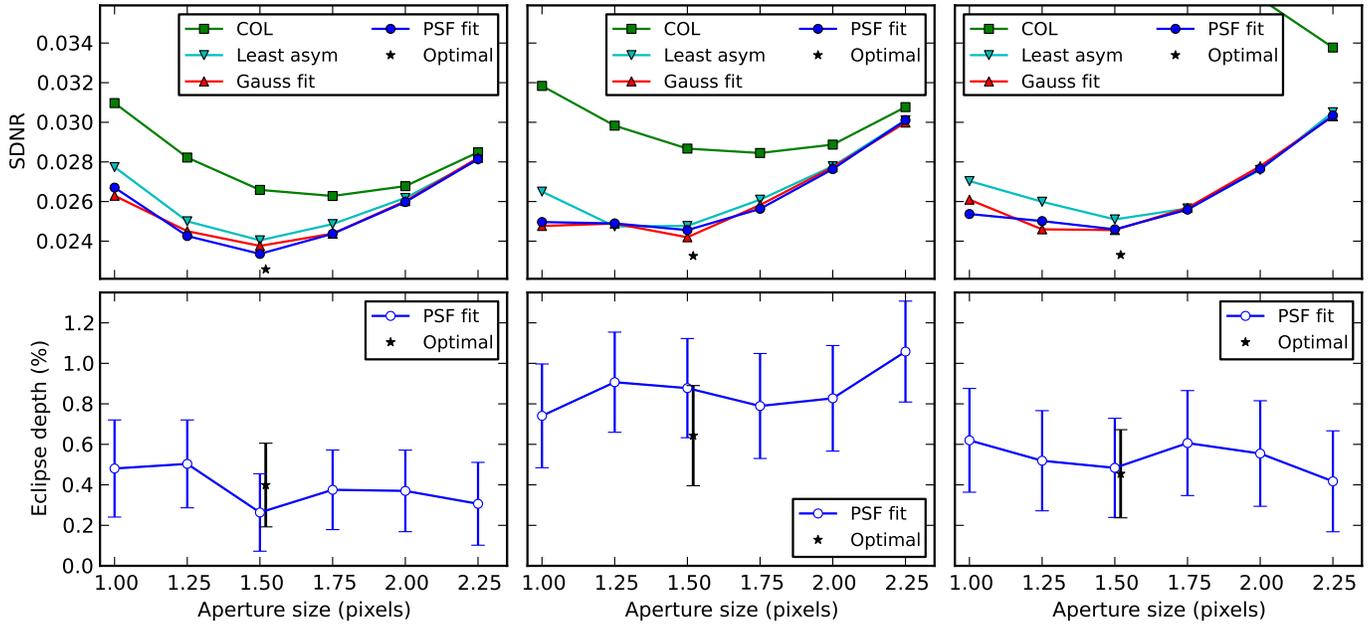}
\caption{{\bf Top:} 16-{\micron} eclipse light-curves SDNR {\vs}
  aperture (from left to right, the first, second, and third visits,
  respectively).    The legend indicates the centering method used;  additionally, the optimal-photometry calculation uses the PSF-fit centering positions but does not involve an aperture.  We plotted the optimal-photometry results next to the best-aperture location (for ease of comparison).  Each curve used the best ramp model from Tables
  \ref{table:tr001bs52ramps}, \ref{table:tr001bs53ramps}, and
  \ref{table:tr001bs51ramps}, respectively. {\bf Bottom:} Eclipse
  depth {\vs} aperture for PSF-fit centering, with the best one
  (optimal photometry) in black.}
\label{fig:sdnrdepthIRS}
\end{figure*}

\subsubsection{IRS - 16-$\mu$m Eclipses}
\label{sec:tr001bs5}

These data come from three consecutive eclipses and present similar
systematics.  The telescope 
cycled among four nodding positions every five acquisitions.  As a
result, each position presented a small flux offset ($\lesssim2\%$).
Since the four nod positions are equally sampled throughout the entire
observation, they should each have the same mean level.
We corrected the flux offset by dividing each frame's flux by the
nodding-position mean flux and multiplying by the overal mean flux,
improving SDNR by $\sim6\%$.
We tested aperture photometry from 1.0 to 5.0 pixels.  In all visits
the SDNR minimum was at an aperture of 1.5 pixels; however,
optimal photometry outperformed aperture photometry
(Figure~\ref{fig:sdnrdepthIRS}).  The second visit provided the
clearest model determination (Table~\ref{table:tr001bs52ramps}).

\begin{table}[ht]
\caption{16-{\microns} eclipse, visit 2 - individual ramp model fits\protect\footnotemark}
\label{table:tr001bs52ramps}
\strut\hfill\begin{tabular}{cccccc}
\hline
\hline
$R(t)$     & Ecl. Depth (\%) & SDNR      & $\Delta$BIC & $p({\cal M}\sb{2}|D)$\\
\hline
linramp    & 0.50(24)        & 0.0233022 & 0.0         & $\cdots$  \\
no-ramp    & 0.40(19)        & 0.0235462 & 4.1         & 0.11 \\
quadramp   & 0.74(28)        & 0.0232539 & 5.3         & 0.06 \\
risingexp  & 0.68(22)        & 0.0232595 & 5.3         & 0.06 \\
\hline
\end{tabular}\hfill\strut
\footnotetext[1]{Fits for PSF-fit centering and optimal photometry.}
\end{table}

At the beginning of the third visit (40 frames, $\sim$ 28 min),
the target position departs from the rest by half a pixel; omitting
the first 40 frames did not improve SDNR.
The linear ramp model minimized BIC (Table
\ref{table:tr001bs53ramps}).  Even though $\Delta$BIC between the
linear and the no-ramp models was small, the no-ramp residuals showed
a linear trend, thus we are confident on having selected the best
model.  The eclipse light curve in this visit is consistent with that
of the second visit.

\begin{table}[ht]
\caption{16-{\microns} eclispe, visit 3 - individual ramp model fits\protect\footnotemark}
\label{table:tr001bs53ramps}
\strut\hfill\begin{tabular}{cccccc}
\hline
\hline
$R(t)$     & Ecl. Depth (\%) & SDNR      & $\Delta$BIC & $p({\cal M}\sb{2}|D)$\\
\hline
linramp    & 0.48(21)        & 0.0233010 & 0.0   & $\cdots$  \\
no-ramp    & 0.24(18)        & 0.0234888 & 1.1   & 0.37 \\
quadramp   & 0.38(22)        & 0.0233004 & 5.8   & 0.05 \\
risingexp  & 0.48(20)        & 0.0233011 & 6.2   & 0.04 \\
\hline
\end{tabular}\hfill\strut
\footnotetext[1]{Fits for PSF-fit centering and optimal photometry.}
\end{table}

The eclipse of the first visit had the lowest S/N of
all. The free parameters in both minimizer and MCMC
easily ran out of bounds towards implausible solutions.
For this reason we determined the best model in a joint fit combining
all three visits.  The events shared the eclipse midpoint,
duration, depth, and ingress/egress times.
We used the best data sets and models from the second and third visits
and tested different ramp models for the first visit.
With this configuration, the linear ramp model minimized the BIC of the
joint fit (Table \ref{table:tr001bs51ramps}).
Here, the target locations in the first two nodding cycles also were
shifted with respect to the rest of the frames.  
Clipping them out improved the SDNR.  Figure
\ref{fig:lightcurves} shows the data and best-fitting light curves and
rms-{\vs}-bin size plot.

\begin{table}[ht]
\caption{16-{\micron} eclipse, visit 1 - ramp model fits\protect\footnotemark}
\label{table:tr001bs51ramps}
\strut\hfill\begin{tabular}{cccccc}
\hline
\hline
$R(t)$     &Ecl. Depth (\%) & SDNR      & $\Delta$BIC      & $p({\cal M}\sb{2}|D)$ \\
tr001bs51  & Joint          & Joint     & Joint            & tr001bs51    \\
\hline
linramp    & 0.35(14)       & 0.0230156 & \phantom{0}0.00  & $\cdots$     \\
quadramp   & 0.32(14)       & 0.0230172 & \phantom{0}7.10  & 0.03         \\
risingexp  & 0.36(11)       & 0.0230152 & \phantom{0}7.31  & 0.02         \\
no-ramp    & 0.33(13)       & 0.0231502 & \phantom{}10.16  & $6\tttt{-3}$ \\
\hline
\end{tabular}\hfill\strut
\footnotetext[1]{Fits for PSF-fit centering and optimal photometry.}
\end{table}

\subsubsection{IRS - 16-$\mu$m Transit}
\label{sec:tr001bs5}

To fit this light curve we used the \citet{MandelAgol2002ApJtransits}
small-planet transit model with a quadratic limb-darkening law.  We
included priors on the model parameters that were poorly constrained
by our data.  We adopted $\cos(i)=0.0\sp{+0.019}\sb{-0.0}$ and
$a/R\sb{\star}=10.52\sp{+0.02}\sb{-0.18}$ from
\citet{TorresEtal2011StellarReanalysis} and the quadratic-limb darkening
coefficients $u\sb{1}=0.284\pm0.061$ and $u\sb{2}=0.21\pm0.12$, which
translate into our model parameters as
$c\sb{2}=u\sb{1}+2u\sb{2}=-0.7\pm0.25$ and
$c\sb{4}=-u\sb{2}=-0.21\pm0.12$ (with $c\sb{1} = c\sb{3} = 0$) from
\citet{WinnEtal2007apjTres1}.  The
midpoint and planet-to-star radius ratio completed the list of
free parameters for the transit model.

\begin{figure}[ht]
\centering
\includegraphics[width=\linewidth, clip]{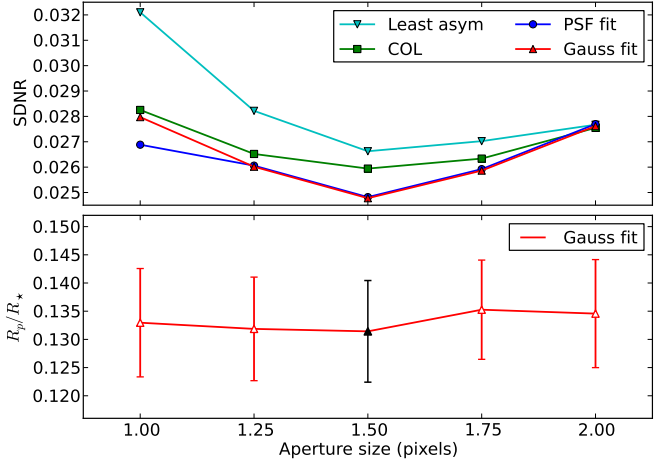}
\caption{ {\bf Top:} 16-{\micron} transit light-curve SDNR {\vs}
  aperture.  All curves used the best ramp model from Table
  \ref{table:tr001bp51ramps}. {\bf Bottom:} Planet-to-star radius
  ratio {\vs} aperture for least asymmetry centering, with the best
  aperture (2.75 pixels) in black.}
\label{fig:sdnrdepth31transit}
\end{figure}

\begin{figure}[thb]
\centering
\includegraphics[width=\linewidth, clip]{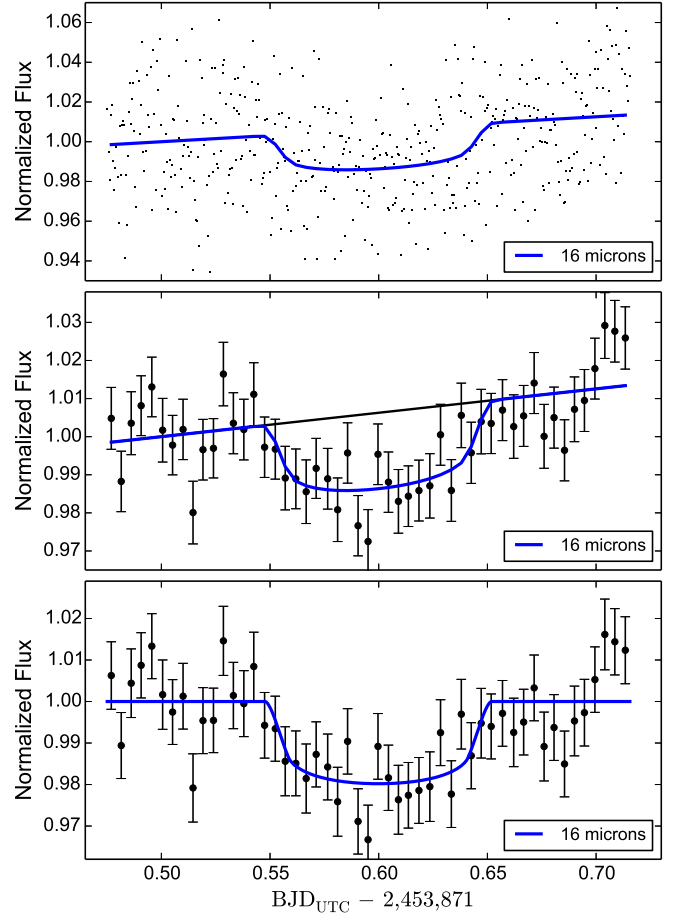}
\caption{Raw (top), binned (middle), and systematics-corrected
  (bottom) normalized TrES-1 transit light curves at 16 {\microns}.  The
  colored curves are the best-fit models.  The black curve is
  the best-fit model excluding the transit component.  The error bars
  are $1\sigma$ uncertainties.}
\label{fig:transitfit}
\end{figure}

We tested aperture photometry between 1 and 2 pixels, finding the SDNR
minimum at 1.5 pixels for the Gaussian-fit centering method
(Figure \ref{fig:sdnrdepth31transit}).  Table
\ref{table:tr001bp51ramps} shows the ramp-model fitting results.  The
linear ramp minimized BIC followed by the quadratic ramp with a 0.33
fractional probability; however, the quadratic fit shows an
unrealistic upward curvature due to high points at the end of the
observation.  Figures \ref{fig:transitfit} and
\ref{fig:lightcurves} show the best fit to the light curve and the
rms-{\vs}-bin size plot, respectively.

\begin{table}[ht]
\caption{16 {\microns} Transit - ramp model fits\protect\footnotemark}
\label{table:tr001bp51ramps}
\strut\hfill\begin{tabular}{cccccc}
\hline
\hline
$R(t)$     & $R\sb{p}/R\sb{\star}$ & SDNR      & $\Delta$BIC & $p({\cal M}\sb{2}|D)$ \\
\hline
linramp    &  0.1314(86)        & 0.0247755  & 0.0         & $\cdots$       \\
quadramp   &  0.1069(224)       & 0.0247118  & 1.4         & 0.33           \\
risingexp  &  0.1314(92)        & 0.0247757  & 6.2         & 0.04           \\
logramp    &  0.1316(81)        & 0.0247768  & 6.3         & 0.04     \\
no-ramp    &  0.1306(89)        & 0.0250938  & 6.9         & 0.03           \\
\hline
\end{tabular}\hfill\strut
\footnotetext[1]{Fits for Gaussian-fit centering and 1.5-pixel aperture photometry.}
\end{table}

\subsubsection{Joint-Fit Analysis} % ::::::::::::::::::::::::::::
\label{sec:joint}

We used the information from all eclipse light curves combined to
perform a final joint-fit analysis.  The simultaneous fit shared a
common eclipse duration, eclipse midpoint and eclipse ingress/egress
time among all light curves.  Additionally, the three IRS eclipses
shared the eclipse-depth parameter.  We further released the duration
prior (which assumed a circular orbit).  We also performed experiments
related to the 3.6 and 4.5 {\micron} datasets.

First, to corroborate our selection of the 3.6 {\micron} model, we
compared the different 3.6 {\micron} models in the joint-fit
configuration both with the shared-midpoint constraint and with
independently-fit midpoints per waveband (Tables
\ref{table:jointramps} and \ref{table:jointmidpoint}).

\begin{table}[ht]
\caption{3.6 {\micron} eclipse models - eclipse-joint fits}
\label{table:jointramps}
\strut\hfill\begin{tabular}{ccccc}
\hline
\hline
$R(t)A(a)$ & $\Delta$BIC   & 3.6 {\micron} Ecl. & Midpoint & Duration  \\
           & 3.6 {\micron} & Depth (\%)         & (phase)  & (phase)   \\  
\hline
\multicolumn{5}{l}{Independently-fit midpoints\footnotemark:}                             \\
$A(a)$     &   0.0         & 0.09(2)    & $\cdots$  & 0.032(1)  \\
quadramp   &   2.3         & 0.16(2)    & $\cdots$  & 0.032(1)  \\
risingexp  &   2.9         & 0.15(2)    & $\cdots$  & 0.032(1)  \\
linramp    &   6.9         & 0.10(2)    & $\cdots$  & 0.032(1)  \\
                                                                       \\
\multicolumn{5}{l}{Shared midpoint:}                                   \\
$A(a)$     &    0.0        & 0.08(2)    & 0.5015(6)        & 0.0328(9) \\
quadramp   &   13.3        & 0.14(3)    & 0.5013(5)        & 0.0331(9) \\
linramp    &   14.2        & 0.08(2)    & 0.5015(6)        & 0.0328(9) \\
risingexp  &   15.0        & 0.12(2)    & 0.5013(5)        & 0.0330(9) \\
\hline
\end{tabular}\hfill\strut
\footnotetext[1]{Midpoint values in Table \ref{table:jointmidpoint}.}
\end{table}

\begin{table}[ht]
\caption{Midpoint per waveband - eclipse-joint fit}
\label{table:jointmidpoint}
\strut\hfill\begin{tabular}{cccccc}
\hline
\hline
$R(t)A(a)$ & 3.6 {\micron} & 4.5 {\micron} & 5.8 {\micron} & 8.0 {\micron} & 16 {\micron} \\
          & (phase)  & (phase)  & (phase)  & (phase)  & (phase)  \\
\hline
$A(a)$    & 0.500(3) & 0.503(1) & 0.502(4) & 0.501(1) & 0.499(3) \\
quadramp  & 0.493(2) & 0.503(1) & 0.502(4) & 0.501(1) & 0.500(4) \\
risingexp & 0.493(1) & 0.503(1) & 0.502(4) & 0.501(1) & 0.500(3) \\
linramp   & 0.491(1) & 0.503(1) & 0.507(4) & 0.501(1) & 0.499(3) \\
\hline
\end{tabular}\hfill\strut
\end{table}

\begin{figure}[thb]
\centering
\includegraphics[width=\linewidth, clip]{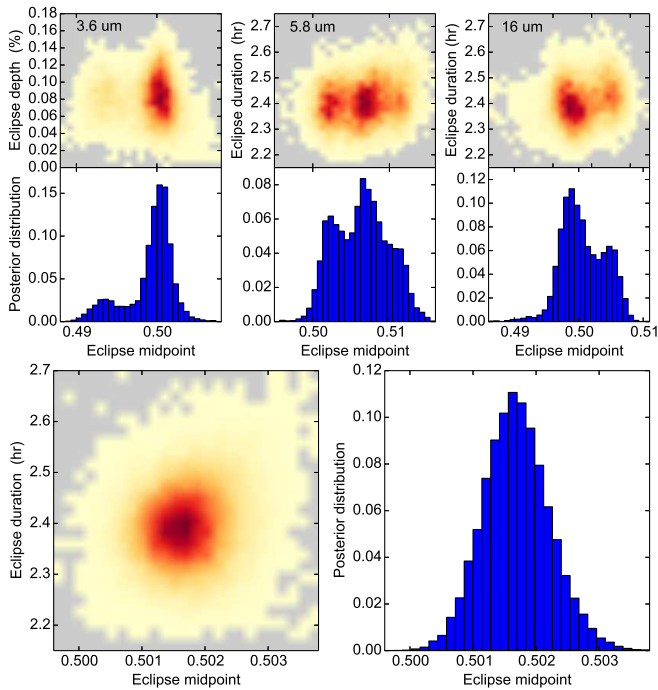}
\caption{Eclipse-midpoint pairwise and marginal posteriors.  {\bf
    Top:} Independently-fit posterior eclipse depth or duration {\em
    vs.} midpoint for (left to right) 3.6, 5.8, and 16.0 {\microns}.
  The multi-modality did not replicate for the eclipse depth (same
  eclipse depth for each of the posterior modes).  {\bf Bottom:}
  Eclipse-duration {\em vs.} midpoint pairwise (left) and midpoint
  marginal (right) posterior distributions for the fit with shared
  midpoint.}
\label{fig:bimodalphase}
\end{figure}

All wavebands other than 3.6 {\microns} agreed with an eclipse
midpoint slightly larger than 0.5.  When we fit the midpoint
separately for each waveband, only the AOR-scale model at 3.6
{\microns} agreed with the other bands' midpoint (note that the 5.8
{\micron} data were obtained simultaneously with the 3.6 {\micron}
data, and should have the same midpoint).  The posterior distributions
also showed midpoint multimodality between these two solutions (Figure
\ref{fig:bimodalphase}).  On the other hand, with a shared midpoint,
the 3.6 {\micron} band assumed the value of the other bands for all
models, with no multimodality.  All but the AOR-scale model showed
time-correlated noise, further supporting it as the best choice.

Second, we investigated the impact of the (potentially corrupted)
4.5~{\micron} data set on the joint-fit values.  Excluding the
4.5~{\micron} event from the joint fit does not significantly alter
the midpoint (phase $0.5011~\pm~0.0006$) nor the duration
($0.0326~\pm~0.013$).  Our final joint fit configuration uses
the AOR-scaling model for the 3.6~{\micron} band, includes the
4.5~{\micron} light curve, and shares the eclipse midpoint (Table
\ref{table:jointfits}).

\subsubsection{4.5  and 8.0 $\mu$m Eclipse Reanalyses} % :::::::::::::::::::::::
\label{sec:comparison}

Our current analysis methods differ considerably from those of nearly
a decade ago, with better centering, subpixel aperture photometry,
BLISS mapping, simultaneous fits across multiple data sets, and
evaluation of multiple models using BIC.
Furthermore, MCMC techniques were not yet prominent in most
exoplanet analyses, among other improvements.
\citet{CharbonneauEtal2005apjTrES1} used two field stars (with similar
magnitudes to TrES-1) as flux calibrators.  They extracted light
curves using aperture photometry with an optimal aperture of 4.0
pixels, based on the rms of the calibrators' flux.  At 4.5 {\microns},
they decorrelated the flux from the telescope pointing, but gave no details.
At 8.0 {\microns}, they fit a third-order
polynomial to the calibrators to estimate the ramp.
Their eclipse model had two free parameters (depth and midpoint),
which they fit by mapping {\chisq} over a phase-space grid.  Table
\ref{table:C05comparison} compares their eclipse depths with ours,
showing a marginal 1$\sigma$ difference at 4.5 {\microns}.  In
both channels our MCMC found larger eclipse-depth uncertainties
compared to those of \citet{CharbonneauEtal2005apjTrES1}, who 
calculated them
from the $\chi\sp{2}$ contour in the phase-space grid.
The introduction of MCMC techniques and the further use of more
efficient algorithms (e.g., differential-evolution MCMC) that converge
faster enabled better error estimates.  In the past, for example, a
highly-correlated posterior prevented the MCMC convergence of some
nuisance (systematics) parameters.  The non-convergence forced one to
fix these parameters to their best-fitting values.  In current
analyses, however, marginalization over nuisance parameters often
leads to larger but more realistic error estimates.

The muxbleed correction was likely less accurately made than required
for atmospheric characterization, given the presence of a visible
muxbleed trail in the background near the star.  We cannot easily
assess either the uncertainty or the systematic offset added by the
muxbleed and its correction, given, e.g., that the peak pixel flux
varies significantly with small image motions.  Our stated 4.5
{\micron} uncertainty contains no additional adjustment for this
unquantified noise source, which makes further use of the 4.5
{\micron} eclipse depth difficult.  However, our minimizer and the
{\chisq} map of \citeauthor{CharbonneauEtal2005apjTrES1} clearly find
the eclipse, so the timing and duration appear less affected than the
depth.  In the analyses below, we include fits both with and without
this dataset.  The large uncertainty found by MCMC limits the 4.5
{\micron} point's influence in the atmospheric fit.

\begin{table}[ht]
\centering
\caption{Eclipse-depth reanalysis}
\label{table:C05comparison}
\strut\hfill
\begin{tabular}{lcc}
\hline
\hline
Eclipse depth (\%)                   & 4.5 {\microns} & 8.0 {\microns} \\
\hline
\citet{CharbonneauEtal2005apjTrES1}  & 0.066(13)      & 0.225(36)      \\
This work                            & 0.094(24)      & 0.213(42)      \\
\hline
\end{tabular}
\hfill\strut
\end{table}

\begin{figure*}[th]
\centering
\includegraphics[width=0.67\linewidth,  clip]{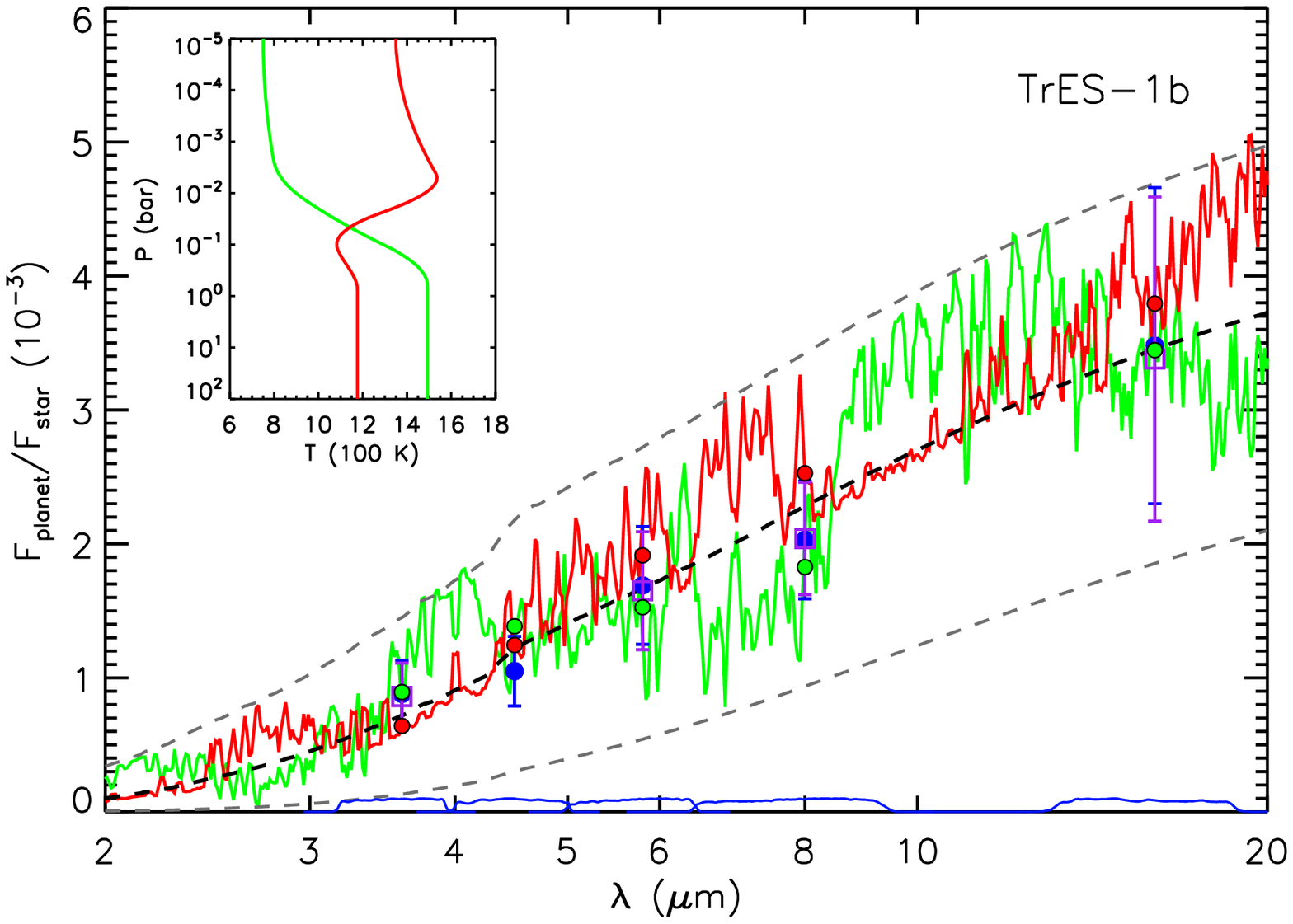}\hfill
\includegraphics[width=0.315\textwidth, clip]{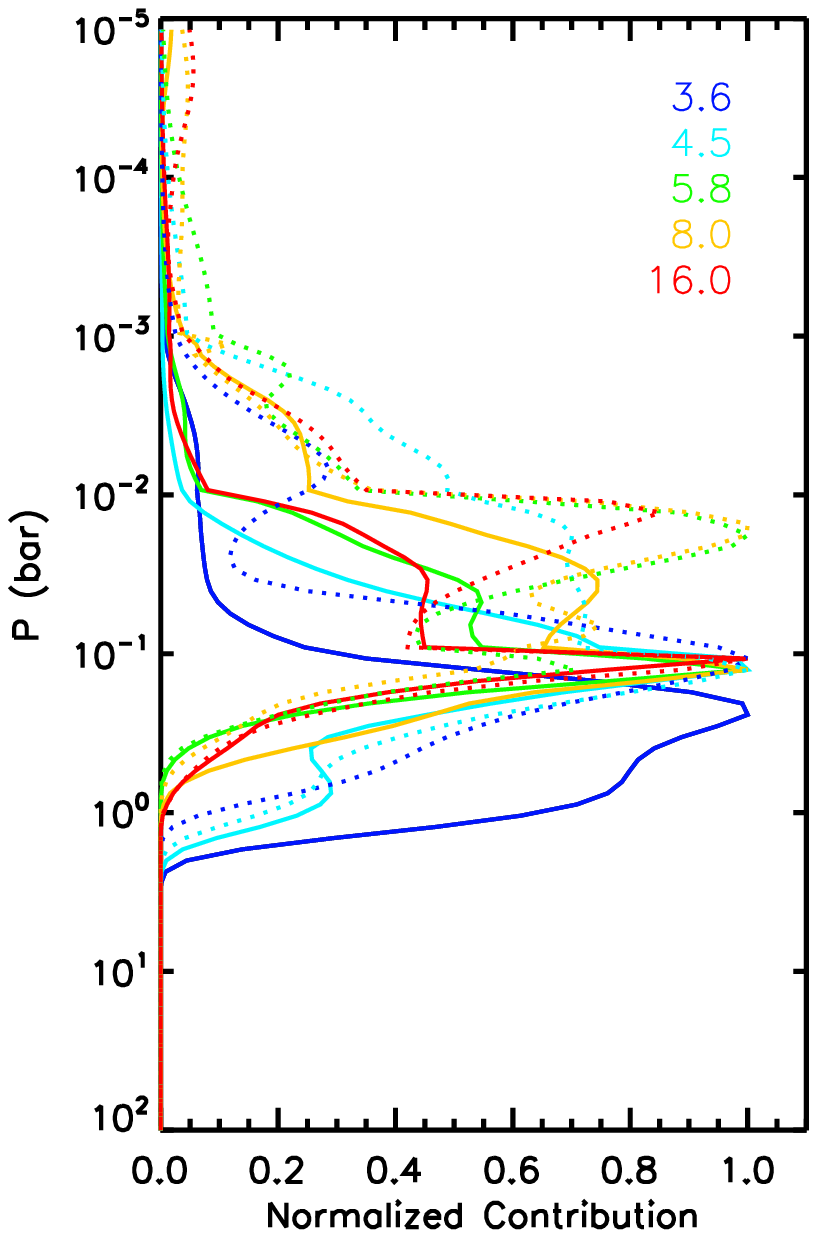}
\caption{{\bf Left:} Dayside atmospheric spectral emission of TrES-1.
  The blue circles and purple squares with error bars are the measured
  eclipse depths (including and excluding the 4.5 {\microns} data
  point, respectively).  The red and green curves show representative
  model spectra with and without thermal inversion (see inset), based
  on the data including the 4.5 {\micron} point.  Results omitting
  this point are similar.  Both models have a solar abundance atomic
  composition and are in chemical equilibrium for the corresponding
  temperature profiles.  The red and green circles give the
  band-integrated (bottom curves) fluxes of the corresponding models,
  for comparison to data.  The dashed lines represent planetary
  blackbody spectra with $T$ = 800, 1200, and 1500 K.  {\bf Right:}
  Normalized contribution functions of the models over each {\Spitzer}
  band (see legend).  The dotted and solid lines are for the models
  with and without thermal inversion, respectively.}
\label{fig:atm}
\end{figure*}

\section{ORBITAL DYNAMICS}
\label{sec:orbit}

As a preliminary analysis, we derived $e\cos(\omega)$ from the eclipse
data alone.
Our seven eclipse midpoint times straddle
phase 0.5.  After subtracting a light-time correction of $2a/c = 39$
seconds, where $a$ is the semimajor axis and $c$ is the speed of
 light, we found an eclipse phase of $0.5015 \pm 0.0006$.  This
implies a marginal non-zero value for $e\cos(\omega)$ of $0.0023 \pm
0.0009$ \citep[under the small-eccentricity approximation,][]{CharbonneauEtal2005apjTrES1}.

It is possible that a non-uniform brightness emission from the planet
can lead to non-zero measured eccentricity \citep{WilliamsEtal2006apj}.  For
example, a hotspot eastward from the substellar point can simulate a
late occultation ingress and egress compared to the uniform-brightness
case.  However, as pointed out by \citep{deWitEtal2012aapFacemap}, to constrain
the planetary brightness distribution requires a higher photometric
precision than what TrES-1 can provide.

Further, using the MCMC routine described by
\citet{CampoEtal2011apjWASP12b}, we fit a Keplerian-orbit model to our
secondary-eclipse midpoints simultaneously with 33 radial-velocity
(Table \ref{table:rv})
and 84 transit data points (Table \ref{table:transit}).
We discarded nine radial-velocity points that were affected by the
Rossiter-McLauglin effect.  We 
were able to constrain $e\cos(\omega)$ to $0.0017 \pm 0.0003$.
Although this 3$\sigma$ result may suggest a non-circular orbit,
when combined with the fit to $e\sin(\omega)$ of $-0.033 \pm 0.025$,
the posterior distribution for the eccentricity only indicates a
marginally eccentric orbit with $e = 0.033\sp{+0.015}\sb{-0.031}$.
Table \ref{table:orbit} summarizes our orbital MCMC results.

\begin{table}[ht]
\vspace{-10pt}
\centering
\caption{\label{table:orbit} MCMC Eccentric Orbital Model}
\begin{tabular}{rc}
\hline
\hline
Parameter                              & Best fitting value                \\
\hline
\math{e\sin\omega}                     & $-0.033$    {\pm}  0.025          \\
\math{e\cos\omega}                     &   0.0017    {\pm}  0.0003         \\
\math{e}                               &   0.033 $\sp{+0.015}\sb{-0.031}$  \\
\math{\omega} (\degree)                & 273      $\sp{+1.4}\sb{-2.8}$     \\
Orbital period (days)                  & 3.0300699   {\pm}  $1\tttt{-7}$   \\
Transit time, $T\sb{0}$ (MJD)\tnm{a}   & 3186.80692  {\pm}  0.00005        \\
RV semiamplitude, $K$ ({\ms})          & 115.5       {\pm}  3.6            \\
system RV, $\gamma$ ({\ms})            & $-3.9$      {\pm}  1.3            \\
Reduced \math{\chi\sp{2}}              & 6.2                               \\
\hline
\end{tabular}
\tablenotetext{1}{MJD = BJD\sb{TDB}$-2,450,000$}
\end{table}

\section{ATMOSPHERE}
\label{sec:atmosphere}

We modeled the day-side emergent spectrum of TrES-1 with the retrieval
method of \citet{MadhusudanSeager2009apjRetrieval} to constrain the
atmospheric properties of the planet.  The code solves the
plane-parallel, line-by-line, radiative transfer equations subjected
to hydrostatic equilibrium, local thermodynamic equilibrium, and
global energy balance.  The code includes the main sources of opacity
for hot Jupiters: molecular absorption from H$\sb{2}$O, CH$\sb{4}$,
CO, and CO$\sb{2}$ \citep[Freedman, personal communication
2009]{FreedmanEtal2008apjsOpacities}, and H$_2$-H$_2$ collision
induced absorption \citep{Borysow2002H2H2}.  We assumed a Kurucz
stellar spectral model \citep{CastelliKurucz2004}.

The model's atmospheric temperature profile and molecular abundances
of H$\sb{2}$O, CO, CH$\sb{4}$, and CO$\sb{2}$ are free parameters,
with the abundance parameters scaling initial profiles that are in
thermochemical equilibrium.  The output spectrum is integrated over
the {\Spitzer} bands and compared to the observed eclipse depths by
means of $\chi\sp{2}$.  An MCMC module supplies millions of parameter
sets to the radiative transfer code to explore the phase space
\citep{MadhusudhanSeager2010apj, MadhusudhanSeager2011apjGJ436b}.

Even though the features of each molecule are specific to certain
wavelengths \citep{MadhusudhanSeager2010apj}, our independent
observations (4 or 5) are less than the number of free parameters
(10), and thus the model fitting is a degenerate problem.  Thus, we
stress that our goal is not to reach a unique solution, but to discard
and/or constrain regions of the parameter phase space given the
observations, as has been done in the past
\citep[e.g.,][]{BarmanEtal2005, Burrows2007HD209,
  KnutsonEtal2008apjHD209, MadhusudanSeager2009apjRetrieval,
  StevensonEtal2010natGJ436b, MadhusudhanEtal2011natWASP12batm}.

Figure \ref{fig:atm} shows the TrES-1 data points and model spectra of
its day-side emission.  An isothermal model can fit the observations
reasonably well, as shown by the black dashed line (blackbody
spectrum with a temperature of 1200 K).  However, given the low S/N of
the data, we cannot rule out non-inverted nor strong thermal-inversion
models (with solar abundance composition in chemical equilibrium), as
both can fit the data equally well (green and red models).
Generally speaking, the data allow for efficient day-night
heat redistribution;  the models shown have maximum possible
heat redistributions of 60\% (non-inversion model) and 40\% (inversion
model).

As shown in Fig.\ \ref{fig:atm}, the data sets with and without the
4.5~{\micron} point are nearly identical.  Combined with the large
error bars (especially at 16 {\microns}), there is no significant difference
between the atmospheric model  results of the two cases. 
Both CO and CO$\sb{2}$ are dominant absorbers at 4.5 {\microns}.
Combined with the 16-{\micron} detection, which is mainly sensitive to
CO$\sb{2}$, the data could constrain the abundances of CO and CO$\sb{2}$.
Unfortunately, the error bar on the 16 {\micron} band is too large to
derive any meaningful constraint.

%% ::::::::::::::::::::::::::::::::::::::::::::::::::::::::::::::::::::
\section{CONCLUSIONS}
\label{sec:conclusions}

We have analyzed all the {\Spitzer} archival data for {\mbox{TrES-1}},
comprising eclipses in five different bands (IRAC and IRS blue
peak-up) and one IRS transit.
There has been tremendous improvement in data-analysis
techniques for {\Spitzer}, and exoplanet light curves in general,
since \citet{CharbonneauEtal2005apjTrES1}, one of the first two
reported exoplanet secondary eclipses.
A careful look at the 4.5 {\micron} data frames revealed pixels
affected by muxbleed that, although corrected by the {\Spitzer}
pipeline, still showed a clear offset output level.  Unable to know
the effect on the eclipse depth and uncertainty, we conducted
subsequent modeling both with and without the 4.5 {\micron} point.
The already-large uncertainty resulted in similar conclusions either
way.  Without adjusting our point for either the systematic or random
effects of the muxbleed correction, the depth and uncertainty at 4.5
{\micron} are both substantially larger than the original analysis.
However, at 8.0 {\microns} (which does not have similar problems) the
eclipse depths are consistent, with our MCMC giving a larger
uncertainty.

Our measured eclipse depths from our joint light-curve fitting (with and without the 4.5 {\micron} point) are
consistent with a nearly-isothermal atmospheric dayside emission at
$\sim1200$ K.  This is consistent with
the expected equilibrium temperature of 1150 K (assuming zero albedo
and efficient energy redistribution).  Furthermore, neither inverted
nor non-inverted atmospheric models can be ruled out, given the low
S/N of the data. Our transit analysis unfortunately does not improve
the estimate of the planet-to-star radius ratio
($R\sb{p}/R\sb{\star}=0.119{\pm}0.009$).  Our comprehensive orbital
analysis of the available eclipse, transit, and radial-velocity data
indicates an eccentricity of $e = 0.034\sp{+0.014}\sb{-0.032}$,
consistent with a circular orbit at the 1$\sigma$-level.  Longitudinal
variations in the planet's emission can induce time offsets in eclipse
light curves, and could mimic non-zero eccentricities
\citep[e.g.,][]{WilliamsEtal2006apj, deWitEtal2012aapFacemap}.
However, the S/N required to lay such constraints are much higher than
that of the TrES-1 eclipse data.

We also described the latest improvements of our POET pipeline.
Optimal photometry provides an alternative to aperture photometry.
We first applied optimal photometry in
\citet{StevensonEtal2010natGJ436b}, but describe it in more detail here.
Furthermore, the Differential-Evolution Markov-chain algorithm poses
an advantage over a Metropolis Random Walk MCMC, since it
automatically tunes the scale and orientation of the proposal
distribution jumps.  This dramatically increases the algorithm's
efficiency, converging nearly ten times faster.  We also now
avoid the need to orthogonalize highly correlated posterior
distributions.  

\acknowledgments

We thank D.\ Charbonneau for sharing the original {\Spitzer} pipeline
data for the 4.5 and 8.0 {\microns} bands and for helpful
discussions. We thank the amateur observers from ETD, including
Alfonso Carre\~no Garcer\'an, Zonalunar Observatory; Ferran Grau, Ca
l'Ou observatory, Sant Mart\'i Sesgueioles; Hana Ku$\check{\rm
  c}$\'akov\'a, Altan Observatory, Czech Republic and Johann Palisa
Observatory and Planetarium, Technical University Ostrava, Czech
Republic; Prof.\ Dr.\ Johannes\ M.\ Ohlert; Christopher De Pree, Agnes
Scott College and SARA; Peter Roomian, College of San Mateo
Observatory; Stan Shadick, Physics and Engineering Physics Dept.,
University of Saskatchewan; Bradley Walter, Meyer Observatory, Central
Texas Astronomical Society.  Thanks Colo Colo for its Camp30n
campaign.  We thank contributors to SciPy, Matplotlib, and the Python
Programming Language, the free and open-source community, the NASA
Astrophysics Data System, and the JPL Solar System Dynamics group for
software and services.  PC is supported by the Fulbright Program for
Foreign Students.  This work is based on observations made with the
{\em Spitzer Space Telescope}, which is operated by the Jet Propulsion
Laboratory, California Institute of Technology under a contract with
NASA.  Support for this work was provided by NASA through an award
issued by JPL/Caltech and through the NASA Science Mission
Directorate's Astrophysics Data Analysis Program, grant NNH12ZDA001N.

\bibliography{TrES1}

\begin{appendices}

\section{Joint Best Fit}
\label{sec:app}
\setcounter{table}{0}
\renewcommand\thetable{\Alph{section}.\arabic{table}}

Table \ref{table:jointfits} summarizes the model setting and results
of the light-curve joint fit.  The midpoint phase parameter was
shared among the IRS eclipse observations.

Table \ref{table:rv} lists the aggregate TrES1 radial-velocity
measurements.

Table \ref{table:transit} lists the aggregate TrES1 transit-midpoint
measurements.

\section{Light-curves Data Sets}

All the light-curve data sets are available in flexible Image
Transport System (FITS) format in a tar.gz package in the electronic
edition.

\begin{turnpage}
\begin{table*}[ht]
\centering
\caption{\label{table:jointfits} Best-Fit Eclipse Light-Curve Parameters}
\begin{tabular}{rcccccccc}
\hline
\hline
Parameter                                       & tr001bs11       & tr001bs21\tnm{a}& tr001bs31       & tr001bs41       & tr001bs51      & tr001bs52      & tr001bs53       & tr001bp51 \\
\hline
% Centering:
Centering algorithm                             & Gauss fit       & Center of Light & Least Asymmetry & Least Asymmetry & PSF fit        & PSF fit        & PSF fit         & Gauss fit   \\
Mean $x$ position (pix)                         & 119.95          & 169.02          & 113.74          & 167.92          & $\cdots$       & $\cdots$       & $\cdots$        & $\cdots$  \\
Mean $y$ position (pix)                         &  82.58          & 118.63          & 83.29           & 117.62          & $\cdots$       & $\cdots$       & $\cdots$        & $\cdots$  \\
$x$-position consistency\tnm{b} (pix)           & 0.014           & 0.019           & 0.021           & 0.019           & 0.038          & 0.036          & 0.040           & 0.045     \\
$y$-position consistency\tnm{b} (pix)           & 0.026           & 0.025           & 0.024           & 0.030           & 0.044          & 0.036          & 0.043           & 0.037     \\
%Photometry:
Optimal/Aperture  photometry size (pix)         & 2.50            & 3.75            & 2.75            & 2.75            & optimal        & optimal        & optimal         &  1.5      \\
Inner sky annulus (pix)                         & 7.0             & 7.0             & 7.0             & 7.0             & 5.0            & 5.0            & 5.0             &  5.0      \\
Outer sky annulus (pix)                         & 15.0            & 15.0            & 15.0            & 15.0            & 10.0           & 10.0           & 10.0            & 10.0      \\
% BLISS Map:
BLISS mapping                                   &  Yes            & Yes             & No              & No              &  No            &  No            &  No             & No        \\
Minimum Points Per Bin                          &   4             &  4              &  $\cdots$       & $\cdots$        &  $\cdots$      &  $\cdots$      &  $\cdots$       & $\cdots$  \\
% Fitting Results:
System flux \math{F\sb{s}} (\micro Jy)          & 33191.4(5.9)    & 21787.0(2.3)    & 14184.5(3.3)    & 8440.7(2.3)     & 1792.3(2.1)    & 1797.2(2.3)    & 1796.6(2.3)     & 857(1.8)  \\
Eclipse depth (\%)                              & 0.083(24)       & 0.094(24)       & 0.162(42)       & 0.213(42)       & 0.33(12)       & 0.33(12)       & 0.33(12)        & $\cdots$   \\
Brightness temperature (K)                      & 1270(110)       & 1126(90)        & 1205(130)       & 1190(130)       & 1270(310)      & 1270(310)      & 1270(310)       & $\cdots$   \\
Eclipse midpoint (orbital phase)                & 0.5015(5)       & 0.5015(5)       & 0.5015(5)       & 0.5015(5)       & 0.5015(5)      & 0.5015(5)      & 0.5015(5)       & $\cdots$   \\
Eclipse/Transit midpoint (MJD\sb{UTC})\tnm{c}   & 3630.7152(16)   & 3309.5283(16)   & 3630.7152(16)   & 3309.5283(16)   & 3873.1204(16)  & 3876.1504(16)  & 3879.1805(16)   & $3871.5998(38)$ \\
Eclipse/Transit midpoint (MJD\sb{TDB})\tnm{c}   & 3630.7159(16)   & 3309.5290(16)   & 3630.7159(16)   & 3309.5290(16)   & 3873.1211(16)  & 3876.1512(16)  & 3879.1812(16)   & $3871.6005(38)$ \\
Eclipse/Transit duration (\math{t\sb{\rm 4-1}}, hrs) & 2.39(7)    & 2.39(7)         & 2.39(7)         & 2.39(7)         & 2.39(7)        & 2.39(7)        & 2.39(7)         & 2.496(33)  \\
Ingress/egress time (\math{t\sb{\rm 2-1}}, hrs) & 0.31(1)         & 0.31(1)         & 0.31(1)         & 0.31(1)         & 0.31(1)        & 0.31(1)        & 0.31(1)         & 0.28(2)    \\
% Transit parameters:
$R\sb{p}/R\sb{\star}$                           & $\cdots$        & $\cdots$        & $\cdots$        & $\cdots$       & $\cdots$       & $\cdots$        & $\cdots$        & 0.1295(95) \\
$\cos(i)$                                       & $\cdots$        & $\cdots$        & $\cdots$        & $\cdots$       & $\cdots$       & $\cdots$        & $\cdots$        & $0.0\sp{+0.000008}\sb{-0.0}$   \\
$a/R\sb{\star}$                                 & $\cdots$        & $\cdots$        & $\cdots$        & $\cdots$       & $\cdots$       & $\cdots$        & $\cdots$        & $10.494\sp{+0.092}\sb{-0.135}$ \\
Limb darkening coefficient, $c2$                & $\cdots$        & $\cdots$        & $\cdots$        & $\cdots$       & $\cdots$       & $\cdots$        & $\cdots$        &   0.75(22)  \\
Limb darkening coefficient, $c2$                & $\cdots$        & $\cdots$        & $\cdots$        & $\cdots$       & $\cdots$       & $\cdots$        & $\cdots$        & $-0.19(11)$ \\
% Nuisance parameters:
Ramp equation ($R(t)$)                          & $A(a)$          & None            & None            & linramp         & linramp        & linramp        & linramp         & linramp    \\
Ramp, linear term ($r\sb{1}$)                   & $\cdots$        & $\cdots$        & $\cdots$        & 0.2455(82)      & 0.182(49)      & 0.151(42)      & 0.118(47)       & 0.063(17)  \\
AOR scaling factor ($A(a\sb{2})$)               & 1.00234(33)     & $\cdots$        & $\cdots$        & $\cdots$        & $\cdots$       & $\cdots$       & $\cdots$        & $\cdots$   \\
% Summary:
Number of free parameters\tnm{d}                & 6               & 5               & 5               & 6               & 6              & 6              & 6               & 8          \\
Total number of frames                          & 3904            & 1518            & 1952            & 1518            & 500            & 500            & 500             & 500        \\
Frames used\tnm{e}                              & 3827            & 1407            & 1763            & 1482            & 460            & 500            & 500             & 492        \\
Rejected frames (\%)                            & 1.97            & 7.31            & 9.68            & 2.37            & 8.0            & 0.0            & 0.0             & 1.6        \\
BIC value                                       & 10103.0         & 10103.0         & 10103.0         & 10103.0         & 10103.0        & 10103.0        & 10103.0         & 533.4      \\
SDNR                                            & 0.0053766       & 0.0026650       & 0.0083273       & 0.0074324       & 0.0223287      & 0.0233603      & 0.0233306       & 0.0248263  \\
Uncertainty scaling factor                      & 0.946           & 1.065           & 1.186           & 0.962           & 0.543          & 0.574          & 0.590           & 0.489      \\
Photon-limited S/N (\%)                         & 99.34           & 89.67           & 74.04           & 63.01           & 8.34           & 7.98           & 7.99            & 10.7       \\
\hline
\end{tabular}
%\begin{minipage}[t]{0.85\linewidth}
\tablenotetext{1}{Data corrupted by muxbleed.}
\tablenotetext{2}{RMS frame-to-frame position difference.}
\tablenotetext{3}{MJD = BJD $-$ 2,450,000.}
\tablenotetext{4}{In the individual fits.}
\tablenotetext{5}{We exclude frames during instrument/telescope
  settling, for insufficient points at a given BLISS bin, and for bad
  pixels in the photometry aperture.}
%\end{minipage}
\end{table*}
\end{turnpage}

\begin{table}[ht]
\vspace{-10pt}
\centering
\caption{Tres-1 Radial-Velocity Data}
\label{table:rv}
\begin{tabular}{cr@{\,{\pm}\,}lc}
\hline
\hline
Date                   & \mctc{RV}               & Reference \\
BJD(TDB) $-$ 2450000.0 & \mctc{(m\,s$\sp{-1}$)}  &           \\
\hline
3191.77001             &  $  60.4 $    &  12.8   & 1   \\
3192.01201             &  $ 115.1 $    &   8.3   & 1   \\
3206.89101             &  $  87.1 $    &  16.0   & 1   \\
3207.92601             &  $  15.8 $    &  10.4   & 1   \\
3208.73001             &  $-113.3 $    &  15.0   & 1   \\
3208.91701             &  $ -98.1 $    &  19.8   & 1   \\
3209.01801             &  $-118.4 $    &  15.3   & 1   \\
3209.73101             &  $  49.8 $    &  15.7   & 1   \\
3237.97926             &  $  68.32$    &  3.66   & 2   \\
3238.83934             &  $-102.23$    &  3.27   & 2   \\
3239.77361             &  $ -24.53$    &  3.25   & 2   \\
3239.88499             &  $  10.00$    &  3.11   & 2   \\
3240.97686             &  $  70.68$    &  3.73   & 2   \\
3907.87017             &  $  18.7 $    &  14.0   & 3   \\
3907.88138             &  $  30.5 $    &  12.5   & 3   \\
3907.89261             &  $  54.6 $    &  12.0   & 3   \\
3907.90383             &  $  24.3 $    &  10.4   & 3   \\
\n3907.91505\sp{a}     &  $  26.4 $    &  11.4   & 3   \\
\n3907.92627\sp{a}     &  $  30.4 $    &  10.9   & 3   \\
\n3907.93749\sp{a}     &  $  22.4 $    &  14.3   & 3   \\
\n3907.94872\sp{a}     &  $   2.9 $    &  11.0   & 3   \\
\n3907.95995\sp{a}     &  $  -7.1 $    &  12.1   & 3   \\
\n3907.97118\sp{a}     &  $ -22.3 $    &  13.3   & 3   \\
\n3907.98240\sp{a}     &  $ -40.5 $    &  13.3   & 3   \\
\n3907.99363\sp{a}     &  $ -39.2 $    &  13.0   & 3   \\
\n3908.00487\sp{a}     &  $ -9.8  $    &  12.2   & 3   \\
\n3908.01609\sp{a}     &  $ -30.5 $    &  13.8   & 3   \\
3908.02731             &  $ -17.7 $    &  13.6   & 3   \\
3908.03853             &  $ -24.7 $    &  12.2   & 3   \\
3908.04977             &  $ -27.5 $    &  11.1   & 3   \\
3908.06099             &  $ -38.2 $    &  13.3   & 3   \\
3908.07222             &  $ -23.7 $    &  11.2   & 3   \\
3908.08344             &  $ -23.0 $    &   9.6   & 3   \\
\hline
\multicolumn{4}{l}{\sp{a} Discarded due to Rossiter-McLaughlin effect.}  \\
\multicolumn{4}{l}{References:
 (1) \citealp{AlonsoEtal2004apjTrES1disc};}           \\
\multicolumn{4}{l}{
 (2) \citealp{LaughlinEtal2005TrES1followup};          
 (3) \citealp{NaritaEtal2007TrES1RLmeasurements}.}     \\
\end{tabular}
\end{table}

\begin{table}[ht]
\centering
\caption{\label{table:transit} TrES-1 Transit Midpoint data}
\begin{tabular}{cll}
\hline
\hline
Midtransit date         &  Error    & Source\tnm{a} \\
BJD(TDB) $-$ 2450000.0  &           &        \\
\hline
6253.23986 & 0.00105 & ETD: Sokov E. N.                                    \\
6198.69642 & 0.00119 & ETD: Roomian P.                                     \\
6198.69600 & 0.00056 & ETD: Shadic S.                                      \\
6177.47937 & 0.00099 & ETD: Emering F.                                     \\
6168.39577 & 0.00042 & ETD: Mravik J., Grnja J.                            \\
6107.79376 & 0.00032 & ETD: Shadic S.                                      \\
6074.46334 & 0.00117 & ETD: Bachschmidt M.                                 \\
6074.46253 & 0.00112 & ETD: Emering F.                                     \\
6071.43377 & 0.00055 & ETD: Carre\~no                                      \\
6071.43165 & 0.00072 & ETD: Gaitan J.                                      \\
6071.43099 & 0.0007  & ETD: Horta F. G.                                    \\
5886.59953 & 0.00048 & ETD: Shadic S.                                      \\
5801.75506 & 0.0004  & ETD: Shadic S.                                      \\
5798.73056 & 0.00049 & ETD: Shadic S.                                      \\
5795.69991 & 0.00053 & ETD: Walter B., Strickland W., Soriano R.           \\
5795.69903 & 0.00064 & ETD: Walter B., Strickland W., Soriano R.           \\
5795.69797 & 0.00055 & ETD: Walter B., Strickland W., Soriano R.           \\
5777.51807 & 0.00056 & ETD: Centenera F.                                   \\
5768.42617 & 0.00042 & ETD: V. Krushevska, Yu. Kuznietsova, M. Andreev     \\
5765.39585 & 0.0004  & ETD: V. Krushevska, Yu. Kuznietsova, M. Andreev     \\
5762.36407 & 0.00037 & ETD: V. Krushevska, Yu. Kuznietsova, M. Andreev     \\
5759.33530 & 0.00049 & ETD: V. Krushevska, Yu. Kuznietsova, M. Andreev     \\
5707.81338 & 0.00093 & ETD: Marlowe H., Makely N., Hutcheson M., DePree C. \\
5680.55402 & 0.00064 & ETD: Sergison D.                                    \\
5671.46700 & 0.00114 & ETD: Ku\v{c}\'akov\'a H.                            \\
5671.46384 & 0.00088 & ETD: Vra\v{s}t\'ak M.                               \\
5671.46382 & 0.0009  & ETD: Br\'at L.                                      \\
5371.48766 & 0.00074 & ETD: Mihel\v{c}i\v{c} M.                            \\
5304.82572 & 0.00084 & ETD: Shadick S.                                     \\
5095.75034 & 0.00075 & ETD: Rozema G.                                      \\
5089.69043 & 0.00109 & ETD: Vander Haagen G.                               \\
5068.48006 & 0.00062 & ETD: Trnka J.                                       \\
5062.42088 & 0.00053 & ETD: Sauer T.                                       \\
5062.42078 & 0.00046 & ETD: Trnka J., Klos M.                              \\
5062.42012 & 0.00046 & ETD: D\v{r}ev\v{e}n\'y R., Kalisch T.               \\
5062.41959 & 0.0006  & ETD: Br\'at L.                                      \\
5062.41797 & 0.00102 & ETD: Ku\v{c}\'akov\'a H., Speil J.                  \\
4998.79649 & 0.0016  & ETD: Garlitz                                        \\
4971.51779 & 0.001   & ETD: Gregorio                                       \\
4968.48904 & 0.00192 & ETD: P\v{r}ib\'ik V.                                \\
4968.48811 & 0.00053 & ETD: Trnka J.                                       \\
4968.48753 & 0.00028 & ETD: Andreev M., Kuznietsova Y., Krushevska V.      \\
4671.54149 & 0.0021  & ETD: Mendez                                         \\
4662.44989 & 0.001   & ETD: Forne                                          \\
4383.68459 & 0.0019  & ETD: Sheridan                                       \\
4380.65579 & 0.0014  & ETD: Sheridan                                       \\
4362.47423 & 0.0002  & \citet{HrudkovaEtal2009TTVsearch}                   \\
4359.44430 & 0.00015 & \citet{HrudkovaEtal2009TTVsearch}                   \\
4356.41416 & 0.0001  & \citet{HrudkovaEtal2009TTVsearch}                   \\
4356.41324 & 0.00096 & ETD: Andreev M., Kuznietsova Y., Krushevska V.      \\
4350.35296 & 0.00036 & ETD: Andreev M., Kuznietsova Y., Krushevska V.      \\
4347.32322 & 0.00028 & ETD: Andreev M., Kuznietsova Y., Krushevska V.      \\
3907.96406 & 0.00034 & \citet{NaritaEtal2007TrES1RLmeasurements}           \\
3901.90372 & 0.00019 & \citet{WinnEtal2007apjTres1}                        \\
3901.90371 & 0.0016  & \citet{NaritaEtal2007TrES1RLmeasurements}           \\
3898.87342 & 0.00014 & \citet{NaritaEtal2007TrES1RLmeasurements}           \\
\hline
\multicolumn{3}{c}{Continued on next page}  \\
\hline
\end{tabular}
\tablenotetext{1}{ETD: amateur transits from the Exoplanet Transit
  Database (\href{http://var2.astro.cz/ETD/index.php}{http://var2.astro.cz/ETD/index.php}) with reported error bars and quality indicator of 3 or better.}
\end{table}

\begin{table}[ht]
\renewcommand\thetable{A.3}
\centering
\caption{TrES-1 Transit Midpoint Data -- Continued from previous page}
\begin{tabular}{cll}
\hline
\hline
Midtransit date         &  Error    & Source\tnm{a} \\
BJD(TDB) $-$ 2450000.0  &           &               \\
\hline
3898.87341 & 0.00014 & \citet{WinnEtal2007apjTres1}              \\
3898.87336 & 0.00008 & \citet{WinnEtal2007apjTres1}              \\
3895.84298 & 0.00015 & \citet{NaritaEtal2007TrES1RLmeasurements} \\
3895.84297 & 0.00018 & \citet{WinnEtal2007apjTres1}              \\
3856.45180 & 0.0005  & ETD: Hentunen                             \\
3650.40752 & 0.00045 & ETD: NYX                                  \\
3550.41568 & 0.0003  & ETD: NYX                                  \\
3547.38470 & 0.0012  & ETD: NYX                                  \\
3256.49887 & 0.00044 & ETD: Ohlert J.                            \\
3253.46852 & 0.00057 & ETD: Pejcha                               \\
3253.46812 & 0.00038 & ETD: Ohlert J.                            \\
3247.40751 & 0.0004  & \citet{CharbonneauEtal2005apjTrES1}       \\
3189.83541 & 0.0019  & \citet{CharbonneauEtal2005apjTrES1}       \\
3186.80626 & 0.00054 & \citet{AlonsoEtal2004apjTrES1disc}        \\
3186.80611 & 0.0003  & \citet{CharbonneauEtal2005apjTrES1}       \\
3183.77521 & 0.0005  & \citet{CharbonneauEtal2005apjTrES1}       \\
3174.68641 & 0.0004  & \citet{CharbonneauEtal2005apjTrES1}       \\
2868.65031 & 0.0022  & \citet{CharbonneauEtal2005apjTrES1}       \\
2856.52861 & 0.0015  & \citet{CharbonneauEtal2005apjTrES1}       \\
2847.43631 & 0.0015  & \citet{CharbonneauEtal2005apjTrES1}       \\
2850.47091 & 0.0016  & \citet{CharbonneauEtal2005apjTrES1}       \\
3171.65231 & 0.0019  & \citet{CharbonneauEtal2005apjTrES1}       \\
3192.86941 & 0.0015  & \citet{CharbonneauEtal2005apjTrES1}       \\
3180.75291 & 0.0010  & \citet{CharbonneauEtal2005apjTrES1}       \\
4356.41492 & 0.00010 & \citet{HrudkovaEtal2009TTVsearch}         \\
4359.44506 & 0.00015 & \citet{HrudkovaEtal2009TTVsearch}         \\
4362.47499 & 0.00020 & \citet{HrudkovaEtal2009TTVsearch}         \\
\hline
\end{tabular}
\tablenotetext{1}{ETD: amateur transits from the Exoplanet Transit
  Database (\href{http://var2.astro.cz/ETD/index.php}{http://var2.astro.cz/ETD/index.php}) with reported error bars and quality indicator of 3 or better.}
\end{table}

\end{appendices}
\end{document}